\documentclass[twocolumn,amsmath,amssymb,10pt,aps]{revtex4-1} 

\usepackage{feynmp}
\usepackage{amsmath}
\usepackage{graphicx}
\usepackage{rotating}
\usepackage{multirow}
\usepackage{latexsym}
\usepackage{textcomp}
\usepackage{verbatim}
\usepackage{color}
\usepackage{pict2e}
\usepackage{bm}
\usepackage{subfigure}
\usepackage{everysel}
\usepackage{keyval}
\usepackage{ragged2e}
\usepackage{dsfont}
\usepackage{amssymb}
\usepackage{enumitem}
\usepackage{pstricks}
\usepackage{mathtools}
\usepackage{hyperref}

\newcommand{\osum}{{%
    \setbox0\hbox{\circ}%
    \rlap{\hbox to \wd0{\hss\sum\hss}}\box0
}}

\begin{document}

\title{Topological Phase Transitions in Line-nodal Superconductors}   


\author{SangEun Han}
\thanks{These authors contributed equally to this work.}
\author{Gil Young Cho}
\thanks{These authors contributed equally to this work.}
\author{Eun-Gook Moon}

\affiliation{Department of Physics, Korea Advanced Institute of Science and Technology, Daejeon 305-701, Korea}

\date{\today}    

\begin{abstract} 
Fathoming interplay between symmetry and topology of many-electron wave-functions has deepened understanding of quantum many body systems, especially after the discovery of topological insulators. Topology of electron wave-functions enforces and protects emergent gapless excitations, and symmetry is intrinsically tied to the topological protection in a certain class. Namely, unless the symmetry is broken, the topological nature is intact. 
We show novel interplay phenomena between symmetry and topology in topological phase transitions associated with line-nodal superconductors.
The interplay may induce an exotic universality class in sharp contrast to that of the phenomenological Landau-Ginzburg theory. Hyper-scaling violation and emergent relativistic scaling are main characteristics, and the interplay even induces unusually large quantum critical region. We propose characteristic experimental signatures around the phase transitions in three spatial dimensions, for example, a linear phase boundary in a temperature-tuning parameter phase-diagram.

\end{abstract}

\maketitle
 
Superconductivity is one of the most intriguing quantum many body effects in condensed matter systems : electrons form Cooper pairs whose Bose-Einstein condensation becomes an impetus of striking characteristics of superconductors (SCs), for example the Meisner effect and zero-resistivity \cite{Tinkham}. The pair formation suppresses gapless fermionic excitation and only the superconducting order parameter becomes important in conventional SCs. But, in the unconventional SCs, fermionic excitation is not fully suppressed generically, so the order parameter and fermions coexist and reveal intriguing unconventional nature \cite{Sigrist, matsuda2006nodal, sachdev_keimer}.

The fermionic excitation in unconventional SCs is often protected and classified by its topological nature. A path (or surface) in momentum space around nodal excitation defines a topological invariant  in terms of the Berry phase (flux) of the Bogoliubov de-Gennes (BdG) Hamiltonian. 
In literature \cite{matsuura2013protected,chiu2014classification,chiu2015classification}, structure of the BdG Hamiltonian has been extensively studied and is applied to weakly correlated systems. Proximity effects between topologically different phases (or defects) have been investigated and experimentally tested, focusing on a search for novel excitation such as Majorana modes \cite{majorana1,majorana2}. 

Among unconventional SCs, we focus on a class whose topological nature is protected by a symmetry. Namely, unless the symmetry is broken, topologically-protected nodal structure is intact. In this class, unwinding of topological invariant and spontaneous symmetry breaking appear concomitantly at quantum critical points, and thus intriguing interplay between symmetry and topology is expected to appear. 
Therefore, topological phase transitions around the class of the unconventional SCs become a perfect venue to investigate the interplay between topology and symmetry. 
%

In 2d, Sachdev and coworkers have investigated this class in the context of d-wave SCs \cite{Vojta1,Vojta2,Vojta3}. They found the universality class of point-node vanishing phase transitions is {that of the Higgs-Yukawa theory}, the theory with relativistic fermions and bosons in 2d. 
 
Richer structure exists in three spatial dimensions (3d). In addition to point-nodes, line-nodes are available in 3d.  
Effective phase space of line-nodal excitation is qualitatively distinct from that of order parameter fluctuation as shown by {codimension} analysis \cite{matsuura2013protected,
chiu2014classification,chiu2015classification}. 
Thus, concomitant appearance of symmetry breaking and topological unwinding in line-nodal SCs has us expect an exotic universality class of the topological transitions.

Another motivation of our work is abundant experimental evidence of line-nodal SCs in various strongly correlated systems in heavy fermions 
\cite{CePtSi1, CePtSi2,CePtSi4,gasparini2010superconducting,UCoGe2} 
and pnictides \cite{iron1,iron2,iron3,iron4,iron5}, for example, 
CePtSi${}_3$, UCoGe, (Ba${}_{1-x}$K${}_x$)Fe$_2$As${}_2$, Ba(Fe${}_{1-x}$Co${}_x$)${}_2$As${}_2$, and FeSe in addition to the $^3$He polar superfluid phase \cite{polar}. 
 Reported line-nodal SCs are often adjacent to another superconducting phase with a different symmetry group.
Due to the symmetry difference, nodal structures of two different SC phases are expected to be different, so they become ideal target systems of this work.

In this work, we investigate quantum phase transitions out of line-nodal SCs where intriguing interplay between topology and symmetry appears. We first provide a general rule to investigate adjacent phases of the line-nodal SCs. Then, phase transitions are investigated by standard mean field analysis, which shows generically continuous phase transitions.  
A novel universality class of the continuous phase transitions is discovered and characterized by {\it hyper-scaling violation} and {\it relativistic scaling} with wide quantum critical region. Its striking experimental consequences are also discussed at the end.    
  
\begin{figure}
\begin{center}
\includegraphics[width=\columnwidth]{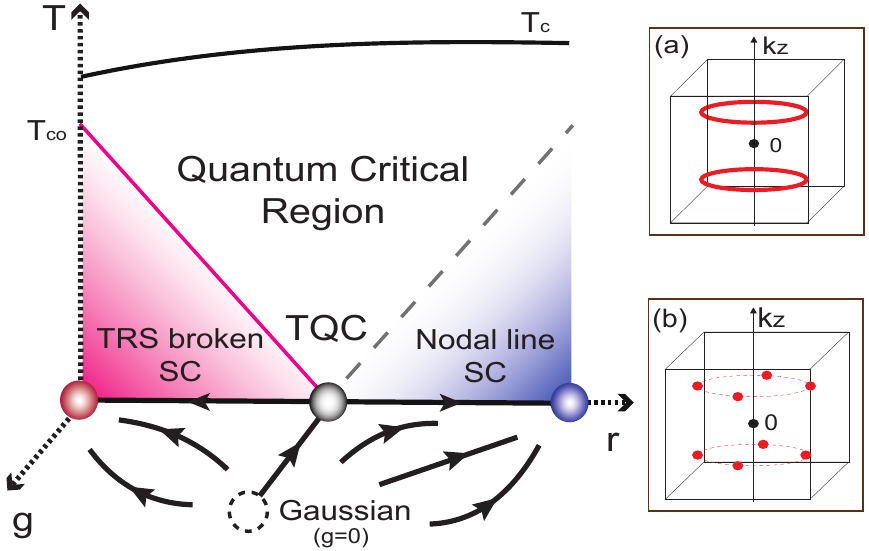}
\caption{Phase Diagram and RG flow.
Three axes are for temperature (T), the tuning parameter (r), and the coupling between order parameter and line-node fermions (g). In r-T plane, critical region is parametrically wider than conventional $\phi^4$ theory's. 
In r-g plane, the RG flow is illustrated by arrows. The ``Gaussian'' fixed point has Laudau MFT's critical exponents due to the upper critical dimension. Once the coupling g turns on, the Gaussian fixed point becomes destabilized and RG flows go into `TQC'. At T=0, the left (red) sphere is for the ordered phase, and the right (blue) sphere is for the disordered phase. $T_c$ plays the high energy cutoff, and $T_{co}$ is for critical temperature of the symmetry breaking order parameter. (a) Nodal lines in momentum space in the symmetric phase are illustrated at $k_z = \pm k_z^* $ in addition to the zero point $\bm{k}= \bm{0}$ (black dot). (b) nodal points in momentum space in a symmetry broken phase (8 nodal points).  
 \label{Fig1}
}
\end{center}
\end{figure}
 
Topological line-nodal SCs protected by a symmetry maintain their nodal structure unless the protecting symmetry is broken. 
Therefore, adjacent symmetry-broken phases may be described by representations of the symmetry. 
For example, the polar phase with line-nodes, A-phase with point-nodes, and nodeless B-phase in liquid $^3$He are described by investigating symmetry representations of SO(3)$_{L} \times $ SO(3)$_{S} \times  $ U(1)$_{\phi}$. 
Below, we take the group $\mathcal{G} = C_{4v} \times \mathcal{T}\times \mathcal{P}$, one of the common lattice groups in line-nodal SC experiments (here $\mathcal{P}$ and $\mathcal{T}$ are for particle-hole and time-reversal symmetries), as a prototype.
Its generalization to other groups is straightforward. 

It is well understood in literature \cite{matsuura2013protected} that the SC model with the symmetry group $\mathcal{G}$
\begin{align}
H_0 = \sum_{\bm{k}} \Psi^{\dagger}_{{\bm k}} \Big( h(\bm{k})\tau^z + \Delta(\bm{k})\tau^x \Big) \Psi_{\bm{k}},  
\label{2:MicroHamiltonian}
\end{align}
has {line-nodes protected by $\mathcal{T}$-symmetry}. 
A four component spinor $\Psi_{\bm k}^{\dagger} = (\psi_{\bm k}^{\dagger}, i\sigma^{y} \psi^{T}_{-{\bm k}})$ where $\psi_{\bm k}^{\dagger} = (c_{\bm{k}, \uparrow}^{*},c_{\bm{k}, \downarrow}^{*})$ is introduced, and the particle-hole (spin) space Pauli matrices $\tau^{x,y,z}$ ($\sigma^{x,y,z}$) are used. The $\tau^z$ term describes a normal state spectrum $h({\bm k}) = \epsilon({\bm k}) - \mu + \alpha \vec{l}({\bm k}) \cdot {\vec{\sigma}}$, and the 
$\tau^x$ term describes a pairing term $\Delta(\bm{k}) = (\Delta_s + \Delta_t {\vec d}({\bm k}) \cdot {\vec{\sigma}})$. 
Energy dispersion $\epsilon({\bm k}) = -2t (\cos(k_x) + \cos(k_y) + \cos(k_z))$ is introduced with spin-orbit coupling strength $\alpha$. The orbital axis of the pairing and spin-orbit terms are identical ${\vec d}({\bm k}) = {\vec l}(\bm{k})=(\sin(k_x),\sin(k_y),0)$ which usually maximizes $T_c$\cite{BrydonTc, FrigeriTc}. The pairing amplitudes $\{ \Delta_s, \Delta_t \}$ are chosen to be real and positive without losing generality because of the $\mathcal{T}$-symmetry. As illustrated in Fig.1 (a), the system exhibits two topological line nodes separated in momentum space protected by the $\mathcal{T}$-symmetry.

It is obvious that $\mathcal{T}$-symmetry breaking superconductivity (the term with ${\tau^y}$) changes nodal structure, so order parameter representations for topological phase transitions can be illustrated as in Table I.  
Group theory analysis guarantees coupling terms between order parameters and fermionic excitation,   
\begin{align}
H_{\psi-\phi} = \sum_{s}  \phi_s \sum_{\bm{k}} \Psi^{\dagger}_{\bm{k}} \mathcal{F}_s({\bm{k}}) M^s \Psi_{\bm{k}}, \nonumber
\end{align} 
where $s$ is for representations (and multiplicity) and $\mathcal{F}_s \, M^s$ are illustrated in Table I. For detail of this classification, see the supplemental material (SM) \textbf{A}. Note that $s = E$ is a two dimensional representation, so the corresponding order parameter ($\phi_{s=E}$) has two components.  
 
\begin{table}[tb!]
\begin{tabular} {c|c|c|c}
\hline
Rep. & Lattice ($\mathcal{F}_s({\bm{k}}) M^s$) &Continuum & \#  \\ \hline
$A_{1}$& $\tau^y$ & $\tau^y$ &  0  \\ \hline
$A_{2}$ & $\sin(k_x) \sin(k_y)(\cos(k_x)- \cos(k_y))\tau^y$ &$\sin(4\theta) \tau^y$ &16  \\ \hline
$B_{1}$ & $(\cos(k_x)- \cos(k_y))\tau^y$ &$\cos(2\theta) \tau^y$ & 8 \\ \hline
$B_{2}$ & $\sin(k_x)\sin(k_y) \tau^y $ &$\sin(2\theta) \tau^y$   & 8 \\ \hline
$E$ & $\sin(k_x)\sin(k_z) \tau^y$, &$\cos(\theta)\tau^y \mu^z$,  &4 \\ 
{}& $\sin(k_y)\sin(k_z) \tau^y$ & $\sin(\theta) \tau^y \mu^z$ & {} \\ \hline
\end{tabular}
\caption{ $\mathcal{C}_{4v}$ representations for topological phase transitions. For simplicity, $\mathcal{T}$ broken and spin-singlet representations are only illustrated. The first column is for representations. The second column is the matrix structure in the Nambu space. The third column is for  continuum representations near nodal lines. The last column is for the numbers of the nodal points in each representation.}
\label{tab:classification}
\end{table}

Two topologically different cases exist. First, momentum independence of $A_1$ representation makes the fermion spectrum gapped completely, so-called $i s$ pairing. 
In the $^3$He context, this phase corresponds to the {weakly $\mathcal{T}$-broken analogue of the B-phase}. Second, the order parameters in $A_{2}$, $B_{1}$, $B_{2}$, and $E$-representations leaves point nodes due to angular dependence. Nodal points appear when $\mathcal{F}_s({\bf k})$ has zeros on line nodes and is in fact Weyl nodes. This phase corresponds to the A-phase in $^3$He.

Armed with understanding of adjacent symmetry broken phases, we consider topological phase transitions. Standard mean field theory (MFT) with on-site interaction $-u(\Psi^{\dagger} \tau^y \Psi)^{2}$ gives a mean field  free energy density of `isotropic' $A_{1}$ representation order parameter ($i$-s pairing),  
\begin{eqnarray}
\mathcal{F}_{MF} (\phi) = (\frac{1}{u} - \frac{1}{u_c}+  T) \phi^2 + k_f |\phi|^3 + \cdots. \label{MFT}
\end{eqnarray}
Coefficients of each term are scaled to be one and $\cdots$ is for higher order terms. Notice that the unusual $|\phi|^3$ term appears whose presence is solely from line-nodal fermions manifested by $k_f$. It also guarantees the phase transition is continuous and makes the usual $\phi^4$ term irrelevant. Furthermore, the order parameter critical exponent becomes significantly different from one of the Landau MFT (which only contains bosonic degrees of freedom),  $\langle \phi \rangle \sim (u-u_c)$ giving $\beta=1$ which already suggests a novel universality class. 

We investigate quantum criticality around the continuous phase transitions. 
For simplicity, we omit the subscript $r$ and introduce one real scalar field $\phi$ to describe the order parameter. Its generalization to the $E$ representation with two scalar fields is straightforward. 

In the phenomenological Landau-Ginzbug theory, order parameter fluctuation near quantum phase transitions may be described by 
\begin{eqnarray}
S_{\phi} = \int_{x, \tau}\frac{1}{2} (\partial_{\tau} \phi)^2+\frac{1}{2}(\nabla \phi)^2 + \frac{r}{2}  \phi^2 + \frac{\lambda}{4 !} \phi^4.
\end{eqnarray}  
Of course, this action is not complete in our systems and necessary to be {supplemented by the corrections from fermions}. Without the coupling between the order parameter and fermions, the critical theory $\mathcal{S}_{\phi}$ with $r=r_c$ is well understood, so-called the $\phi^4$ theory : in 3d, it is at the upper-critical dimension. Thus, the Landau MFT works well upto logarithmic correction and hyper-scaling is satisfied.   
Below, we show that the coupling to the fermions significantly changes low energy physics and induces a novel universality class.

The total action with fermions is   
\begin{eqnarray}
S_{c} = S_{\phi} + S_{\psi}, \quad 
S_{\psi} = \int_{x, \tau} \Psi^{\dagger}(\partial_{\tau} + \mathcal{H}_0) \Psi + g\int_{\tau} H_{\psi-\phi}.  \nonumber
\end{eqnarray}
A coupling constant $g$ characterizes strength of the coupling between the fermions and bosons, and the Hamiltonian density $\mathcal{H}_0$  is introduced ($ H_0 = \int_x \Psi^{\dagger}\mathcal{H}_0 \Psi $).   

Near the phase transitions, low- energy and momentum degrees of freedom become important, so we only need the low-energy continuum theory of the BdG Hamiltonian Eq.\eqref{2:MicroHamiltonian} near nodes and obtain
\begin{align}
\mathcal{H}_0(\bm{k}) &\approx v_z \delta k_z \mu^z  \tau^z + v_\perp \delta k_\perp \tau^x, 
\label{2:Continuum}
\end{align}
where the momentum is $\bm{k} = ((k_f + \delta k_\perp) \cos(\theta_k),  (k_f + \delta k_\perp) \sin(\theta_k), k^{*}_z \mu^z + \delta k_z )$. Here $\mu^z = \pm1$ represents  ``which-line-node" index and the effective parameters $\{ v_z, v_\perp\}$ are the functions of the microscopic parameters. 
The low energy fermion spectrum  (say, $\mu^z=+1$) without the fermion-boson coupling is 
\begin{eqnarray}
\epsilon_{0} (\delta k_z, \delta k_{\perp}, \theta_k) = \pm \sqrt{(v_z\delta k_z)^2+(v_{\perp}\delta k_{\perp})^2 }.  
\end{eqnarray}
One parameter, the angle $0 \le\theta_k \le 2 \pi $, characterizes zero energy states, so a nodal line exists in momentum space.  

Density of states near zero energy vanishes linearly in $\epsilon$, $\mathcal{D}_f(\epsilon) \sim k_f  |\epsilon|$ in a sharp contrast to ones of Fermi surfaces ($\sim \epsilon^0$), nodal points  ($\sim \epsilon^2$), and order parameters ($\sim \epsilon^2$).  
It is clear that the phase space of the nodal-line fermion excitation is different from {that of fluctuation of the order parameter}.  
Such difference in the phase spaces of the bosons and fermions is a consequence of the codimension mismatch.  

The coupling term is also  written in terms of the low-energy degrees of freedoms 
\begin{align}
g \int_{x} H_{\psi-\phi} &\approx  g \int_{\bm{k}, \omega, \bm{q},\Omega}  \phi_{\bm{q}, \Omega} \mathcal{F}(\theta_{\bm{k}}) \Psi^{\dagger}_{\bm{k}+\bm{q},\omega+\Omega} \mathcal{M} \Psi_{\bm{k},\omega},  \nonumber
\end{align}
so-called the Yukawa coupling. 
We use Shankar's decomposition of fermion operators around the line node, $\Psi_{{\bf k}} \approx \Psi(\delta k_z, \delta k_{\perp}, \theta_k ; \mu^z)$.

The standard large-$N_f$ analysis is performed by introducing $N_f$-copies of fermion flavors coupled to the boson $\phi$. 
The lowest order boson self-energy $\Sigma_b (\Omega, \bm{q})$ can be obtained by the usual bubble diagram.
For $A_{1}$ representation, the boson self-energy is  
\begin{align}
\Sigma_b (\Omega, \bm{q}) = & N_f g^2 \int_{\bm{k}, \omega}  \text{Tr}\Big[\tau^y G_{f,0} (\omega, \bm{k})\tau^y G_{f,0} (\omega+\Omega, \bm{q}+ \bm{k}) \Big], \nonumber 
\end{align}
where $G_{f,0}^{-1} (\omega, \bm{k}) = -i\omega + \mathcal{H}_0(\bm{k})$ is the bare fermion propagator.
Notice that the integration is over fermionic momentum and frequency, thus main contribution comes from the line-nodal fermions. Basically, the momentum integration can be replaced with energy integration with $\mathcal{D}_f(\epsilon) \sim k_f |\epsilon|$. 
The integration is straightforward (see SM \textbf{C.1}) and we find
\begin{eqnarray}
\delta \Sigma_b (\Omega, \bm{q})  = \mathcal{C}(k_f N_f) \sqrt{\Omega^2 + v^2_z q^2_z + v_\perp^2 q_\perp^2 }\,\, el[ \rho(\Omega, \bm{q})], \nonumber 
\end{eqnarray} 
with $\delta \Sigma_b \equiv \Sigma_b (\Omega, \bm{q}) - \Sigma_b (0, \bm{0})$ and $\mathcal{C}=\frac{g^2 }{4\pi v_\perp v_z}$. The complete elliptic integral $el[x]$ and variable $\rho(\Omega, \bm{q})=1/(1+\frac{   \Omega^2 + v^2_z q^2_z   }{v_\perp^2 q_\perp^2} )$ are used. 
The elliptic integral is well-defined in all range of momentum and frequency, thus as the lowest order approximation, one can treat the integral as a constant since $1 \le el[x]<2$. 

Two remarks follow. First, the linear dependence in momentum and frequency can be understood by power-counting with the linear fermionic density of states.  
Second, the boson propagator contains the factor $N_f k_f$. Thus, one can understand the large-$N_f$ analysis as an expansion with $\frac{1}{N_f k_f}$ factor. The presence of $k_f$ already suggests suppression of infrared divergences in loop-calculations (see below).

The modified boson action is 
\begin{eqnarray}
S^{eff}_{\phi} &=& \int_{\bm{q},\Omega} \frac{|\phi_{ \bm{q}, \Omega}|^2}{2} \Big( \tilde{r} +\bm{q}^2 + \Omega^2  + \delta \Sigma_b (\Omega, \bm{q}))  \Big)   +\cdots,   \nonumber 
\end{eqnarray}
with $\tilde{r} = r+ \Sigma_b(0, \bm{0})$. 
The self-energy manifestly dominates over the bare terms at long wavelength, thus the bare terms may be ignored near the critical point ($\tilde{r}=0$) and the boson propagator becomes $G_b (\Omega, \bm{q}) \rightarrow \delta \Sigma_b(\Omega, \bm{q})^{-1} $.

The back-reaction of the bosons to the fermions is obtained by the fermion self-energy,    
\begin{align}
\Sigma_f (\omega, \bm{k}) &= g^2  \int_{\Omega, \bm{q}}  \tau^y G_f (\omega+\Omega, \bm{k}+\bm{q})\tau^y G_b (\Omega, \bm{q}). \nonumber
\end{align}  
Straightforward calculation shows the corrections to the parameters of the bare fermion action Eq.\eqref{2:Continuum} has the following structure, 
\begin{align}
\frac{\delta \Sigma_f (\omega, \bm{k})}{\delta \epsilon^{a}} \propto ~~ \frac{1}{N_f \,k_f} \times (\Lambda - \mu), \nonumber
\end{align}
where $\epsilon^{a} = (\omega, v_\perp \delta k_\perp, v_z k_z )$, and $\Lambda$ and $\mu$ are the ultraviolet (UV) and infrared (IR) cutoffs. $k_f$ is the largest momentum scale, $k_f \gg \Lambda \gg \mu$ in this work. 
The same cutoff dependence in the vertex correction is found (omitted here and see SM \textbf{C} for details). 

Two remarks follow. First, the momentum integration captures order parameter fluctuation, so it may be replaced with energy integration with $\sim \epsilon^2$ density of states. Next, the cutoff dependence is a result of the large-$N_f$ expansion with $k_f$ as discussed before. The absence of the infrared divergence indicates perturbation theory works well. Thus, fermions and bosons become basically decoupled at low energy. In renormalization group sense, this indicates the vertex operator is irrelevant at low energy.    
For other representations, the corresponding angle dependent functions $\mathcal{F}(\bm{k}) \mathcal{ M } $ appear in the integrands (see SM \textbf{C.1} for details) which does not modify divergence structure.

The critical theory associated with topological line-nodal SCs is   
\begin{eqnarray}
\mathcal{S}_{\phi}^c 
&=& \int_{\Omega, \bm{q}}  N_f k_f \sqrt{\Omega^2 + v^2_z q^2_z + v_\perp^2 q_\perp^2 } \, \mathcal{R}( \rho(\Omega, \bm{q}))  \frac{|\phi|^2}{2} , \label{QC}
\end{eqnarray}
setting $\tilde{r}=0$. $\mathcal{R}(\rho(\Omega, \bm{q}))$ is an order one non-zero positive well-defined function to characterize representations (see SM \textbf{C.1}). 
Therefore, critical exponents do not depend on $\mathcal{R}(x)$.  
We omit the  $\phi^4$ term which is justified below. 


Let us list striking characteristics of our critical theory. First, the damping term , $ k_f|\Omega|$, at $\bm{q}=0$ exists.  The presence of the damping term appears due to the absence of the Ward identity in our systems in a sharp contrast to  line-nodal normal semimetal with the Coulomb interaction. Its form is the same as the Hertz-Millis theory of antiferromagnetic transitions, but momentum depedence is also linear, so the dynamic critical exponent is relativistic ($z=1$).

Moreover,
the anomalous dimension of the order parameter is large ($\eta_{\phi} =1$), so the scaling dimension of the order parameter  is $[\phi]= \frac{d+z-2+\eta_{\phi}}{2} = \frac{3}{2}$. 
This is completely different from one of the Landau theory ($\phi^4$ theory) at the upper critical dimension ($d=3$ with $z=1$). Due to the large anomalous dimension, the correlation length behaves $  \xi^{-1} \sim |r-r_c| $, so $\nu=1$.
Also, the anomalous dimension makes the $\phi^4$ coupling irrelevant, $[\lambda] <0 $. So our critical theory is stable which becomes a sanity-check of the MFT in Eqn.\eqref{MFT}.

The susceptibility exponent is  $\gamma=1$, and the Fisher equality is satisfied $(2-\eta_{\phi})\nu = \gamma$.
Basically decoupled fermions and bosons contribute to specific  heat independently, $C_v \sim a_f T^2 +a_b T^3$. The first term is from line-nodal fermions, and the second term is from order parameter fluctuations with $d/z=3$ (see SM \textbf{D}).  

The hyper-scaling is {\it violated} even in 3d. If not, one would get the order parameter critical exponent, $\beta$ ($\langle \phi \rangle \sim (r_c - r)^{\beta}$) by the scaling relation, $\beta = \frac{(d+z-2+\eta)\nu}{2} = \frac{3}{2}$. 
But, we already observe $\beta=1$ in our MFT in Eqn. \eqref{MFT}, and also the perturbative calculation in our critical theory gives (see SM \textbf{E}) 
\begin{eqnarray}
\tilde{r} + \Sigma_b(0,\bm{0} ; T) -\Sigma_b(0,\bm{0} ; T=0) \sim \tilde{r} + T, \nonumber  
\end{eqnarray}
giving the critical temperature scaling, $T_{co} \sim |r_c -r| = \tilde{r}$ which gives qualitatively wider quantum critical region than one of the Landau MFT, $T_{co, L} \sim \sqrt{r_c - r}$. 
The hyper-scaling violation indicates the Yukawa coupling is {\it dangerously} irrelevant.
In Table II, we compare our critical theory with other critical theories in 3d \cite{sachdev, srednicki, moon1,savary,herbut, hertz, millis} in terms of critical exponents and hyper-scaling applicability.

Remark that our low energy theory has a larger symmetry than one of the original system, namely $U(1)$ rotational symmetry not the original $C_{4v}$. Thus, $k_f$ is independent of the angle $\theta_{\bm{k}}$. 
This is an artifact of the linearization approximation, but it is not difficult to see the universality class is not modified by inclusion of symmetry breaking terms down to $C_{4v}$ unless singular fermion spectrum  such as nesting appears.

This is because the codimension mismatch is the key of linear dependence of momentum and frequency in the boson self-energy with the presence of $k_f$ and the absence of IR divergence in the fermion self-energy. 
Thus, all critical exponents are the same as ones of Eqn.\eqref{QC}. This is also consistent with previous literature on quantum criticality
\cite{sung-sik2015mixing,huh2015nodal}. 
We also explicitly show the linear dependence without the linearized fermion dispersion approximation in supplementary information.  

\begin{table}[tb!]
\begin{tabular} {|c|c|c|c|c|c|c|}
\hline \hline
QCP in $3d$ & $z$  & ~~$\nu$~~  & ~~$\beta$~~ & ~~$\gamma$~~ & ~~$\eta$~~ & ~~HS~~  \\ \hline \hline
$\phi^4$ theory\cite{sachdev} & $1$ & $\frac{1}{2}$ & $\frac{1}{2}$ & 1 & 0 & O \\ \hline 
Higgs-Yukawa\cite{sachdev, srednicki} & $1$ & $\frac{1}{2}$ & $\frac{1}{2}$ & 1& 0 & O \\ \hline 
QBT-QCP\cite{moon1,savary} & $2$ & $1$ & $2$ & 1 & 1 & O \\ \hline 
 \hline
Hertz-Millis\cite{hertz,millis} & $2$ or $3$  & $\frac{1}{2}$ & $\frac{1}{2}$ & 1 & 0 & X \\ \hline 
Nodal line QCP & $1$    & $1$ & $1$ & 1& 1 & X \\ \hline \hline
\end{tabular}
\caption{Critical theories of QCP in three spatial dimensions ($d=3$). The first raw includes critical exponents ($\Omega \sim q^z$, $\xi^{-1} \sim |r-r_c|^\nu$, $\chi_{\phi} \sim |r-r_c|^{-\gamma}$, and $[\phi] = \frac{d+z-2+\eta}{2}$ ). `HS' is for hyper-scaling.
Both Higgs-Yukawa and $\phi^4$ theory are at the upper critical dimension, so the exponents are ones of the Landau MFT. 
  Both quadratic band touching quantum critical point (QBT-QCP) and Nodal line QCP have wider quantum critical region $\nu=1$ with large anomalous dimension $\eta=1$ obtained by large-$N_f$ analysis. 
}
\end{table}

We now discuss experimental implication of our results. 
First, our results provide additional smoking gun signature of line-nodal SCs. Namely, the linear phase boundary $T_{co} \sim (r_c -r)$, from hyper-scaling violation, between two different SCs identifies the presence of line nodes. Interestingly, some experiments in heavy fermions, for example UCoGe, suggested that a phase boundary between two different SCs is linear 
\cite{gasparini2010superconducting,UCoGe2}and one of SCs at least has line-nodes though further thorough investigation is necessary. 
 
Our analysis indicates that continuous quantum phase transitions associated with line-nodal SCs have a linear phase boundary.  
We argue its converse statement also works. 
Quantum criticality without line-nodal SCs have at most point-nodal fermion excitation. Then, codimensions of order parameter fluctuation and fermion excitation are the same. Therefore, the Yukawa term and $\phi^4$ terms would be (marginally) irrelevant as usual. Thus, we expect quantum criticality without line-nodes would have Landau MFT critical exponents with logarithmic corrections. Detailed discussion on this point will appear in another place.

Furthermore, direct measurement of critical exponents is possible. In particular, 
the discussed fluctuation of the $\mathcal{T}$-breaking order parameters has been extensively studied in a context of $p$-wave SCs in high temperature SCs\cite{Raman1, Raman2, Raman3, Neutron1, Neutron2, Sigrist1}.
Following the literature \cite{muon1, Sigrist1, Sigrist2}, one can investigate a concrete way to measure the fluctuation, namely, the spin polarized muon scattering. From our critical exponents, we obtain the change in the distribution $\delta \sigma$ of internal magnetic fields relative to the $\mathcal{T}$-symmetric phase is $\delta \sigma (r,T) \propto \langle\phi(r,T)\rangle$. Then, our scaling analysis gives  $\delta \sigma (r,T) \propto (r_c-r)\, \mathcal{F}(\frac{T}{r-r_c})$ with a scaling function $\mathcal{F}$. 
Thus, the $\mathcal{T}$-breaking signal is qualitatively different from that of the Landau MFT result $\delta \sigma_{L} (r_c,T=0) \propto \sqrt{r_c-r }$, which manifestly shows consequences of the hyperscaling violation. 

In conclusion, we have described topological phase transitions associated with line-nodal SCs where topology and symmetry reveal intriguing interplay phenomena. We find quantum criticality naturally appears and its universality class of the transitions shows novel characteristics such as emergent relativistic scaling, hyperscaling violation, and unusually wide quantum critical region. { Our results can also be applied to topological phase transitions out of normal nodal ring semi-metals naturally if chemical potential is fixed to be zero.} Future theoretical studies should include more comprehensive treatment of perturbations of critical points such as finite temperature and magnetic field effects. Concrete connection with experiments especially in heavy fermion systems would be also desirable.

\begin{acknowledgments}
It is great pleasure to acknowledge valuable discussion with H. Choi, Y. Huh, and Y. B. Kim.  
E.-G. Moon especially thanks S.-S. Lee for discussion about UV/IR mixing and Y. Huh and Y. B. Kim for previous  collaboration. This work was supported by the Brain Korea 21 PLUS Project of Korea Government and KAIST start-up funding. 
\end{acknowledgments}
%

\bibliographystyle{apsrev}
\bibliography{ref}

\onecolumngrid
\clearpage
\begin{center}
\textbf{\large Supplemental Material for ``Topological Phase Transitions in Line-nodal Superconductors"}
\end{center}
\begin{center}
{SangEun Han, Gil Young Cho, and Eun-Gook Moon}\\
\emph{Department of Physics, Korea Advanced Institute of Science and Technology, Daejeon 305-701, Korea}
\end{center}
\setcounter{equation}{0}
\setcounter{figure}{0}
\setcounter{table}{0}
\setcounter{page}{1}

\appendix

\section{Order Parameters and Nodal Structure of $\mathcal{T}$-broken Phases}
In this supplemental material, we present the detailed derivation of the order parameters (in Table 1.) of the lattice model Eq.(1) whose symmetry group is $C_{4v} \times \mathcal{T} \times \mathcal{P}$ symmetry (here $\mathcal{T}$ and $\mathcal{P}$ represent the time-reversal symmetry and the particle-hole symmetry) and the nodal structures of $\mathcal{T}$-broken phases. We also discuss the polar phase of He$^{3}$, i.e., the nodal line phase experimentally found in He$^{3}$, and its proximate phases.  

\subsection{Derivation of Order Parameters} 
Here we start with the lattice Hamiltonian Eq. (1) of the maintext,  
\begin{align}
H_0 = \sum_{\bm{k}} \Psi^{\dagger}_{{\bm k}} \Big( h(\bm{k})\tilde{\tau}^z + \Delta(\bm{k})\tilde{\tau}^x \Big) \Psi_{\bm{k}},  
\end{align}
with $h({\bm k}) = \epsilon({\bm k}) - \mu + \alpha \vec{l}({\bm k}) \cdot {\vec{\sigma}}$, where $\epsilon({\bm k}) = -2t (\cos(k_x) + \cos(k_y) + \cos(k_z))$ and $\vec{l}(\bm{k}) = (\sin(k_x), \sin(k_y),0)$. The pairing is given by $\Delta(\bm{k}) = \Delta_s + \Delta_t \vec{l}(\bm{k}) \cdot \vec{\sigma}$. The full cubic lattice symmetry is broken by $\vec{l}({\bm k}) = (\sin(k_x), \sin(k_y), 0)$ down to $C_{4v}$ with the $C_4$ rotation in $xy$-plane. 

We first demonstrate the existence of the line nodes by employeeing the basis which diagonalizes $\vec{l}(\bm{k}) \cdot \vec{\sigma} = \pm |\vec{l}(\bm{k})|$, i.e., the helicity basis. In this basis, the Hamiltonian can be block-diagonalized, depending on the signs of the eigenvalues of $\vec{l}(\bm{k}) \cdot \vec{\sigma}$, i.e., $\vec{l}(\bm{k}) \cdot \vec{\sigma} \to \sigma |\vec{l}(\bm{k})|, \sigma = \pm1$, to write  
\begin{align}
H_0 = \sum_{\bm{k}}\Big( \Psi_{+, \bm{k}}^{\dagger} H_{+}(\bm{k})\Psi_{+,\bm{k}} + \Psi_{-, \bm{k}}^{\dagger} H_{-}(\bm{k})\Psi_{-, \bm{k}} \Big), 
\label{Helicity_Hamiltonian}
\end{align}
where 
\begin{align}
H_{\sigma} (\bm{k}) = \begin{pmatrix}
\epsilon({\bm k}) - \mu + \sigma \alpha |\vec{l}(\bm{k})| & \Delta_s + \sigma \Delta_t |\vec{l}(\bm{k})| \\ 
\Delta_s + \sigma \Delta_t |\vec{l}(\bm{k})| & -\Big(\epsilon({\bm k}) - \mu + \sigma \alpha |\vec{l}(\bm{k})|  \Big)  \\ 
\end{pmatrix}= \Big(\epsilon({\bm k}) - \mu + \sigma \alpha |\vec{l}(\bm{k})|  \Big) \tau^z + \Big(\Delta_s + \sigma \Delta_t |\vec{l}(\bm{k})| \Big) \tau^x, 
\end{align}
where we have introduced the Pauli matrix $\tau^{x,y,z}$ acting on the two-component Nambu spinor $\chi_{\sigma, \bm{k}}$. On writing the Hamiltonian into this form, we can easily calculate the BdG spectrum 
\begin{align}
E_{\sigma}(\bm{k}) = \pm \sqrt{\Big(\epsilon({\bm k}) - \mu + \sigma \alpha |\vec{l}(\bm{k})|  \Big)^2 + \Big(\Delta_s + \sigma \Delta_t |\vec{l}(\bm{k})| \Big)^2 }. 
\end{align}
Because $\alpha >0$, $\Delta_s >0$ and $\Delta_t >0$, $\sigma= +1$ is fully gapped and $\sigma = -1$ is nodal if $2 \Delta_t >\Delta_s$ (see below). The position of the zero-energy manifold of the BdG fermion, i.e., line nodes, are identified by 
\begin{align}
&\Delta_s = \Delta_t |\vec{l}(\bm{k})|, \nonumber\\ 
& \epsilon({\bm k}) - \mu - \alpha |\vec{l}(\bm{k})| = 0. 
\label{ConditionLineNode}
\end{align}

From the above discussion, it is apparent that the terms $\propto \tilde{\tau}^y \to \tau^y$ will gap out the nodes. To see this clearly, we first imagine to add $\delta H =  \phi \sum_{\bm{k}}\mathcal{F}(\bm{k})\Psi^{\dagger}_{\bm{k}}\tilde{\tau}^y \Psi_{\bm{k}}$ to $H_0$ to find 
\begin{align}
H_0 + \delta H = \sum_{\bm{k}} \Psi^{\dagger}_{{\bm k}} \Big( h(\bm{k})\tilde{\tau}^z + \Delta(\bm{k})\tilde{\tau}^x  + \phi \mathcal{F}(\bm{k})\tilde{\tau}^y \Big) \Psi_{\bm{k}}.   
\end{align}
By proceeding to the helicity basis again, we find that 
\begin{align}
H_{\sigma} (\bm{k}) \to \Big(\epsilon({\bm k}) - \mu + \sigma \alpha |\vec{l}(\bm{k})|  \Big) \tau^z + \Big(\Delta_s + \sigma \Delta_t |\vec{l}(\bm{k})| \Big) \tau^x + \phi \mathcal{F}(\bm{k})\tau^y, 
\end{align}
whose BdG spectrum is given by 
\begin{align}
E_{\sigma}(\bm{k}) = \pm \sqrt{\Big(\epsilon({\bm k}) - \mu + \sigma \alpha |\vec{l}(\bm{k})|  \Big)^2 + \Big(\Delta_s + \sigma \Delta_t |\vec{l}(\bm{k})| \Big)^2 + \phi^2 \mathcal{F}^2(\bm{k}) }.
\end{align}
We are particularly interested in $\sigma = -1$ which is of the lowest energy, and $E_{-} (\bm{k})$ can be zero if 
\begin{align}
&\Delta_s = \Delta_t |\vec{l}(\bm{k})|, \nonumber\\ 
& \epsilon({\bm k}) - \mu - \alpha |\vec{l}(\bm{k})| = 0, \nonumber\\ 
& \mathcal{F}(\bm{k}) = 0,   
\end{align}
which are more stringent conditions than Eq.\eqref{ConditionLineNode}. Hence the term $\propto \tilde{\tau}^y$ lifts the line nodes to the full gap for $\mathcal{F}(\bm{k})$ being nonzero constant on the line node, or the point nodes for $\mathcal{F} (\bm{k})$ having the zeros on the line node. 

Hence we classify the possible mass term according to the symmetry $C_{4v} \times \mathcal{T}\times \mathcal{P}$ for the lattice model. For the classification, it is instructive to write out the mass term 
\begin{align}
 \delta H = \phi \sum_{\bm{k}}\mathcal{F}(\bm{k})\Psi^{\dagger}_{\bm{k}}\tilde{\tau}^y \Psi_{\bm{k}} =  \phi \times \sum_{\bm{k}}\Big( i \mathcal{F}(\bm{k}) \psi_{\bm{k}}^{\dagger} (i\sigma^y) \psi^{*}_{-\bm{k}}  + h.c.,\Big), 
\end{align}
which is the \textit{imaginary component} of the singlet pairing between the electrons (remember $\psi_{\bm{k}} = (c_{\bm{k}, \uparrow}, c_{\bm{k}, \downarrow})^T$). Hence we immediately notice that it is time-reversal odd, i.e., $\delta H$ breaks $\mathcal{T}$-symmetry as expected (otherwise, the line node is protected and stable). Furthermore, it is part of the pairing and thus, by definition, is particle-hole symmetric. 

Secondly, it is the singlet pairing between the electrons. This implies that 
\begin{align}
g \in C_{4v}: \sum_{\bm{k}} \mathcal{F}(\bm{k}) \psi_{\bm{k}}^{\dagger} (i\sigma^y) \psi^{*}_{-\bm{k}} \to \sum_{\bm{k}} \mathcal{F}(g^{-1}[\bm{k}]) \psi_{\bm{k}}^{\dagger} (i\sigma^y) \psi^{*}_{-\bm{k}}, 
\end{align}
in which $g^{-1}[\bm{k}]$ is the map of $\bm{k}$ under $g^{-1}$ with $g\in C_{4v}$ (because the symmetry $C_{4v}$ is a unitary symmetry). Thus the form factor $\mathcal{F}(\bm{k})$ solely determines the representation class of the order parameters. Now given this information, it is straightforward to classify the mass terms (or order parameters). \newline {}\newline
\textbf{1}.\textit{$A_{1}$ representation}: $\mathcal{F} (\bm{k}) = \text{const}$. ($is$- pairing)\newline
\textbf{2}.\textit{$A_{2}$ representation}: $\mathcal{F} (\bm{k}) = \sin(k_x) \sin(k_y)(\cos(k_x)- \cos(k_y)) $. ($ig$-pairing)\newline
\textbf{3}.\textit{$B_{1}$ representation}: $\mathcal{F} (\bm{k})= \cos(k_x)- \cos(k_y)$. ($id_{x^2 - y^2}$-pairing)\newline 
\textbf{4}.\textit{$B_{2}$ representation}: $\mathcal{F} (\bm{k}) = \sin(k_x)\sin(k_y) $. ($id_{xy}$-pairing)\newline
\textbf{5}.\textit{$E$ representations}: $\mathcal{F}(\bm{k}) =\sin(k_x)\sin(k_z)$, or $\mathcal{F}(\bm{k}) = \sin(k_y)\sin(k_z)$.($id_{xz}$- and $id_{yz}$- pairings) \newline{}\newline
This is the set of the order parameters present in table 1 of the maintext. 

\subsection{Nodal Structure of $\mathcal{T}$-broken Phases}\label{Continuum}
We now present the detailed nodal structure of the $\mathcal{T}$-broken phases. To investigate the nodal structure, it is beneficial to proceed to the low-energy continuum limit of the lattice Hamiltonian Eq. (1) in the maintext and the $\mathcal{T}$-breaking order parameters in table 1 of the maintext. 

To project to the low-energy limit, we first ignore the $\sigma = +1$ band and take only the $\sigma=-1$ band in Eq. \eqref{Helicity_Hamiltonian} supplemented by the approximations $\epsilon({\bm k}) = -2t (\cos(k_x) + \cos(k_y) + \cos(k_z)) \to \frac{k^2}{2m}$ with $m = \frac{1}{2t}$ and $\vec{l}({\bm k}) = (\sin(k_x), \sin(k_y), 0) \approx (k_x, k_y, 0) = \bm{k}_\perp$. Then it is straightforward to demonstrate that the $\sigma=-1$ band of the BdG Hamiltonian of Eq. \eqref{Helicity_Hamiltonian} becomes, 
\begin{align}
H \approx v_z \delta k_z \tau^z \mu^z + \delta k_\perp (v_\perp \tau^x + \zeta \tau^z),   
\end{align}
in which $\tau^\alpha$ is acting on the particle-hole basis (in this $\sigma = (-1)$-band), $\mu^z$ is acting on the ``valley" index, i.e., $\mu^z = +1$ for the node at $k_z = k^{*}_z$ and $\mu^z = -1$ for the node at $k_z = - k^{*}_z$. Here 
\begin{align}
v_z = \frac{k^{*}_z}{m},~v_\perp = -\Delta_t, \text{ and }~\zeta = \frac{k_f}{m}-\alpha. 
\end{align}

Thus the low-energy Hamiltonian is given by 
\begin{align}
H\approx \sum_{\delta k_\perp, \delta k_z}\int \frac{d\theta}{2\pi}~\Psi(\delta k_z, \delta k_\perp, \theta)^{\dagger} \Big((v_z \delta k_z \mu^z + \zeta \delta k_\perp) \tau^z + v_\perp \delta k_\perp \tau^x \Big) \Psi(\delta k_z, \delta k_\perp, \theta),
\end{align}
in which the momentum of the quasiparticle is given by 
\begin{align}
\bm{k} = (k^{*}_z \mu^z + \delta k_z, (k_f + \delta k_\perp) \cos(\theta),  (k_f + \delta k_\perp) \sin(\theta) ), 
\end{align} 
i.e., we have moved from the cartesian coordinate to the polar coordinate. 

To investigate the nodal structure of $\mathcal{T}$-broken phases, we next project the coupling between the order parameters and BdG fermions $\delta H = \phi \sum_{\bm{k}}\mathcal{F}(\bm{k})\Psi^{\dagger}_{\bm{k}}\tilde{\tau}^y \Psi_{\bm{k}}$ to the $\sigma=-1$ band. The projection can be effectively done through :
\begin{align}
&\sin(k_x) \to k_x \propto \cos(\theta_{\bm{k}}),\nonumber\\ 
&\sin(k_y) \to k_y \propto \sin(\theta_{\bm{k}}), \nonumber \\ 
&\cos(k_x) - \cos(k_y) \to k_x^2 - k_y^2 \propto \cos(2\theta_{\bm{k}}), \nonumber\\ 
&\sin(k_z) \to \mu^z.  
\end{align} 
With this in hand, we can write out the mass terms in terms of the low-energy fermions, 
\begin{align}
\delta H = \phi \sum_{\delta k_\perp, \delta k_z}\int \frac{d\theta}{2\pi}~\Psi(\delta k_z, \delta k_\perp, \theta)^{\dagger} \Big(\mathcal{F}(\theta) \tau^y + \mathcal{G}(\theta) \tau^y\mu^z\Big) \Psi(\delta k_z, \delta k_\perp, \theta), 
\end{align}  
where we have \newline{}\newline
\textbf{1}.\textit{$A_{1}$ representation}: $\mathcal{F} (\theta) = 1$. $\mathcal{G}=0$.\newline
\textbf{2}.\textit{$A_{2}$ representation}: $\mathcal{F} (\theta) = \sin(4\theta)$. $\mathcal{G}=0$.\newline
\textbf{3}.\textit{$B_{1}$ representation}: $\mathcal{F} (\theta)= \cos(2\theta)$. $\mathcal{G}=0$.\newline 
\textbf{4}.\textit{$B_{2}$ representation}: $\mathcal{F} (\theta)= \sin(2\theta)$. $\mathcal{G}=0$.\newline
\textbf{5}.\textit{$E$ representations}: $\mathcal{F}(\theta) =0$. $\mathcal{G}(\theta) = \cos(\theta)$ or $\mathcal{G}(\theta) = \sin(\theta)$ (two-dimensional representation). \newline{}\newline
With these in hand, we can now investigate the nodal structures of each phase. Specifically, we will show the existence of the Weyl nodes for $A_{2}$, $B_{1}$, $B_{2}$, and $E$ representations, and that of the full gap for $A_{1}$ representation. We will mainly consider the Hamiltonian for the one-dimensional representation cases  
\begin{align}
H\approx \int_{\bm{k}}\Psi^{\dagger}_{\bm{k}} \Big((v_z \delta k_z \mu^z + \zeta \delta k_\perp) \tau^z + v_\perp \delta k_\perp \tau^x + \phi \mathcal{F}(\theta_{\bm{k}})\tau^y \Big) \Psi_{\bm{k}}, 
\end{align}
but it is straightforward to generalize to the two-dimensional representation $E$. To see the structure clearly, we first transform $v_z \delta k_z \mu^z + \zeta \delta k_\perp \to v_z \delta k_z \mu^z$ by translating $k_z \to k_z - \zeta \delta k_\perp \mu^z/v_z$. Then we have 
\begin{align}
H\approx \int_{\bm{k}}\Psi^{\dagger}_{\bm{k}} \Big(v_z \delta k_z \mu^z \tau^z + v_\perp \delta k_\perp \tau^x + \phi \mathcal{F}(\theta_{\bm{k}})\tau^y \Big) \Psi_{\bm{k}}.  
\end{align}
Below we consider only $A_1$ and $B_2$ representations but the consideration below can be easily generalized to the other representations. 

\subsubsection{$A_{1}$-representation}
We show the full gap of $A_{1}$ representation, we simply need to diagonalize the Hamiltonian 
\begin{align}
H\approx \int_{\bm{k}}\Psi^{\dagger}_{\bm{k}} \Big(v_z \delta k_z \mu^z \tau^z + v_\perp \delta k_\perp \tau^x + \phi \tau^y \Big) \Psi_{\bm{k}}, 
\end{align}
and find 
\begin{align}
E(\bm{k}) = \pm \sqrt{v_z^2 \delta k_z^2 + v_\perp^2 \delta k_\perp^2 + \phi^2}.  
\end{align}
It is clear that as far as $\phi \neq 0$, the spectrum is fully gapped. 

\subsubsection{$B_{2}$-representation and others}
We now keep the dependence on $\mathcal{F}(\theta)$ here. By diagonalizing 
\begin{align}
H\approx \int_{\bm{k}}\Psi^{\dagger}_{\bm{k}} \Big(v_z \delta k_z \mu^z \tau^z + v_\perp \delta k_\perp \tau^x + \phi \mathcal{F}(\theta_{\bm{k}})\tau^y \Big) \Psi_{\bm{k}}, 
\end{align}
we have 
\begin{align}
E(\bm{k}) = \pm \sqrt{v_z^2 \delta k_z^2 + v_\perp^2 \delta k_\perp^2 + \phi^2 \mathcal{F}^2(\theta_{\bm{k}})}.  
\end{align}
The spectrum is gapless where the form factor $\mathcal{F}(\theta)$ vanishes. Other points on the ring such that $\mathcal{F} \neq 0$ will be gapped out. The point nodes are in fact Weyl point nodes which corresponds to the hedgehogs in momentum space. To demonstrate this explicitly, we choose $B_{2}$-representation as an example and expand the mean-field Hamiltonian near the point node at $\theta = 0$ and $\mu^z = 1$ for simplicity. Near this point, the fermionic BdG Hamiltonian can be expanded 
\begin{align}
H = \int_{\bm{k}} \Psi^{\dagger} \Big( v_z k_z \tau^z + v_\perp k_x \tau^x + \frac{2  \phi  k_y}{k_f}\tau^y \Big)\Psi,
\end{align}
in which $\delta k_\perp = k_x$ and $\theta = \frac{k_y}{k_f}$ at the vicinity of $\theta =0$. This is the Hamiltonian for the topological Weyl fermions with the winding number is $+1$. The analysis can be generalized to the other point nodes in the other representations.

\subsection{Line-nodal $p_z$-paired Phase and Proximate $\mathcal{T}$-broken Phases}
Here we discuss the line nodal $p_z$-paired phase which may arise from the liquid He${}^3$. This is so-called the polar phase. The normal state is described by $\xi(\bm{k}) = \frac{\bm{k}^2}{2m}-\mu$. Note the absence of the spin-orbit coupling. We concentrate on a particular line-nodal paired state here, but it can be easily generalized to any line-nodal $p$-wave superconducting state. The pairing state that we are interested in is given by the pairing 
\begin{align}
\Delta_t = \langle c^{\dagger}_{\bm{k}, \alpha} [\vec{d}(\bm{k})\cdot \vec{\sigma}(i\sigma^y)]^{\alpha,\beta}c_{-\bm{k}, \beta}^{\dagger} \rangle,   
\end{align}
in which the orbital axis of the triplet pairing is given by $\vec{d}(\bm{k}) = (0,0, k_z)$. 

For this paired state, we can use the Nambu basis $\Psi_{\bm{k}} = (c_{\bm{k}, \uparrow}, c^{\dagger}_{\bm{k}, \downarrow})$ to write out the BdG Hamiltonian
\begin{align}
H = \sum_{\bm{k}} \Psi^{\dagger}(\bm{k})
\begin{pmatrix}
\xi(\bm{k}) & \Delta_t k_z \\ 
\Delta_t k_z & -\xi(\bm{k})  \\ 
\end{pmatrix}
\Psi(\bm{k}).  
\end{align} 
It is easy to confirm that this paired state has the symmetry group $C_{4v} \times \mathcal{T} \times \mathcal{P}$.  

There is a line node at 
\begin{align}
k_z = 0, ~ \text{and}~ k_f = |(k_x, k_y, 0)| = \sqrt{2m\mu}, 
\end{align}
which is protected by $\mathcal{T}$-symmetry.  Furthermore, by expanding the Hamiltonian near the node, we obtain the low-energy theory  
\begin{align}
H_{\bm{k}} = \frac{k_f}{m}\delta k_{\perp}\tau^z -\Delta_t \delta k_z \tau^x 
\end{align}
renaming the variables $\frac{k_f}{m} \to v_\perp$ and $-\Delta_t \to v_z $, we arrive at the low-energy Hamiltonian  
\begin{align}
H_{\bm{k}} = v_\perp \delta k_{\perp} \tau^z  +v_z \delta k_z \tau^x. 
\end{align}
As in the noncentrosymmetric SC case, we now need to classify the mass terms. To investiate the mass terms, we first identify the symmetry actions on the low-energy BdG fermions.

\textbf{1. Time-reversal symmetry} 
\begin{align}
\mathcal{T} : \Psi(\delta k_z, \delta k_\perp, \theta) \to i\tau^y \Psi^{*}(\delta k_z, \delta k_\perp, \theta)
\end{align}

\textbf{2. Particle-hole symmetry}
\begin{align}
\mathcal{P} : \Psi(\delta k_z, \delta k_\perp, \theta) \to \Psi^{*}(\delta k_z, \delta k_\perp, \theta)
\end{align}

\textbf{3. $C_4$ rotation $(x,y,z) \to (-y,x, z)$}
\begin{align}
C_4 : \Psi(\delta k_z, \delta k_\perp, \theta) \to  \Psi(\delta k_z, \delta k_\perp, \theta + \pi/2)
\end{align} 

\textbf{4. $M_y$ mirror $(x,y,z) \to (x,-y,z)$}
\begin{align}
M_y : \Psi(\delta k_z, \delta k_\perp, \theta) \to \Psi(\delta k_z, \delta k_\perp, -\theta)
\end{align} 

With the symmetry actions in hand, we now classify the order parameters according to the lattice symmetry $C_{4v}$. The mass terms which introduce gap on the line node are obviously of the form  
\begin{align}
\delta H = \phi  \int_{\bm{k}}\mathcal{F}(\theta) \Psi^{\dagger}(\delta k_z, \delta k_\perp, \theta) \tau^y \Psi(\delta k_z, \delta k_\perp, \theta) ,
\end{align}
which breaks the $\mathcal{T}$-symmetry (to open up the gap at the node) and $\mathcal{F}(\theta)$ determines which representation class that the order parameters will belong to. 

As in the non-centrosymmetric superconductor, we can easily classify $\mathcal{F}(\theta)$ to find 
\begin{align}
&A_{1}: \mathcal{F}(\theta) = 1. \nonumber\\ 
&A_{2}: \mathcal{F}(\theta) = \sin(4\theta). \nonumber\\
&B_{1}: \mathcal{F}(\theta) = \cos(2\theta). \nonumber\\ 
&B_{2}: \mathcal{F}(\theta) = \sin(2\theta), \nonumber\\ 
&E: \mathcal{F}(\theta) = \sin(\theta), ~\text{or}~ \mathcal{F}(\theta) = \cos(\theta). 
\end{align}
The order parameters defines the phase: \newline
\textbf{1}:\textit{$A_{1}$}: fully gapped, isotropic phase. This is $p_z + is$ phase. This is \textit{spectrally} equivalent to He$^{3}$ B phase in that it is fully gapped in bulk. \newline
\textbf{2}:\textit{$A_{2}$}: partially gapped, eight Weyl nodes. This is $p_z + ig_{xy}$ phase. \newline
\textbf{3}:\textit{$B_{1}$}: partially gapped, four Weyl nodes. This is $p_z + id_{x^2-y^2}$ phase. \newline
\textbf{4}:\textit{$B_{2}$}: partially gapped, four Weyl nodes. This is $p_z + i d_{xy}$ phase. \newline
\textbf{5}:\textit{$E$}: partially gapped, two Weyl nodes. This is $p_z + ip_x$ or $p_z + ip_y$ phase. \newline 

Furthermore, it is trivial to see that the nature of the $\mathcal{T}$-broken phases as well as the low-energy physics here are identical to those of the non-centrosymmetric SC case. 

\section{Mean Field Analysis of $\mathcal{T}$-breaking Phase Transition}
In this supplemental material, we will perform the mean field analysis of the $\mathcal{T}$-breaking phase transition. We explicitly illustrate the calculation for the one-dimensional representations but it is easy to generalize to the two-dimensional $E$ representation. 

We start with the Hamiltonian
\begin{align}
H=\sum_{\bm{k}}\Psi_{\bm{k}}^{\dagger}H_{\bm{k}}\Psi_{\bm{k}}-\frac{u}{2}( \sum_{\bm{k}}\Psi_{\bm{k}}^{\dagger}\mathcal{F}(\theta_{\bm{k}})\tau_{y}\Psi_{\bm{k}} )^{2}, 
\end{align}
where $H_{k} = v_z \delta k_z \mu^z \tau^z + v_\perp \delta k_\perp \tau^x$ is the low-energy Hamiltonian for the BdG fermion on the nodal line. 

Performing the standard Hubbard-Stratanovich technique, we find
\begin{align}
H_{\text{MF}}=\sum_{\bm{k}}\Psi_{\bm{k}}^{\dagger}(H_{\bm{k}}-\phi\mathcal{F}(\theta_{k})\tau_{y})\Psi_{\bm{k}}+\frac{\phi^{2}}{2u},
\end{align}
where
\begin{align}
\phi=u\langle\Psi_{\bm{k}}^{\dagger}\mathcal{F}(\theta_{\bm{k}})\tau_{y}\Psi_{\bm{k}}\rangle.
\end{align}
The free energy is
\begin{align}
\notag\mathcal{F}_{\text{MF}}(T,\phi)=&-\frac{T}{V}\ln(\text{tr}(e^{-\beta H_{\text{MF}}}))=-\frac{T}{V}\sum_{\bm{k}}\sum_{n}\left(\ln\beta(-i\omega_{n}+E_{\bm{k}}(\phi))+\ln\beta(-i\omega-E_{\bm{k}}(\phi))\right)+\frac{\phi^{2}}{2u}\\
\notag=&-\frac{T}{V}\sum_{\bm{k}}\left( \ln(1+e^{-\beta E_{\bm{k}}(\phi)}) +\ln(1+e^{\beta E_{\bm{k}}(\phi)}) \right)+\frac{\phi^{2}}{2u}\\
\notag=&-\frac{T}{V}\sum_{\bm{k}}\left(\beta E_{\bm{k}}(\phi)+ 2\ln(1+e^{-\beta E_{\bm{k}}(\phi)})\right)+\frac{\phi^{2}}{2u}\\
=&-\frac{1}{V}\sum_{\bm{k}}E_{\bm{k}}(\phi)-\frac{2T}{V}\sum_{\bm{k}}\ln(1+e^{-\beta E_{\bm{k}}(\phi)})+\frac{\phi^{2}}{2u}, 
\end{align}
where $E_{\bm{k}}(\phi)$ is
\begin{align}
E_{\bm{k}}(\phi)=\sqrt{v_{z}^{2}k_{z}^{2}+v_{\perp}^{2}k_{\perp}^{2}+\phi^{2}\mathcal{F}(\theta_{\bm{k}})^{2}},
\end{align}
and the momentum summation becomes
\begin{align}
\sum_{\bm{k}}\rightarrow ck_{f}\int dk_{\perp}dk_{z}d\theta_{k} \sum_{\mu^{z} = \pm1} = \int_{\bm{k}}\sum_{\mu^{z} = \pm1}, 
\end{align}
where $c= \frac{1}{(2\pi)^3}$. The free energy variation due to the order parameter $\phi$ is
\begin{align}
\delta \mathcal{F}_{\text{MF}}(\phi)=&\mathcal{F}_{\text{MF}}(\phi)-\mathcal{F}_{\text{MF}}(0)=\frac{1}{V}\sum_{\bm{k}}(E_{\bm{k}}(0)-E_{\bm{k}}(\phi))-\frac{2T}{V}\sum_{\bm{k}}\ln\left(\frac{1+e^{-\beta E_{\bm{k}}(\phi)}}{1+e^{-\beta E_{\bm{k}}(0)}}\right)+\frac{\phi^{2}}{2u}.
\end{align}
At $T=0$, expanding in term of $\phi$, and determine the critical strength $u_{c}$,
\begin{align}
-\sum_{\bm{k}}\frac{\left(\mathcal{F}(\theta_{k})\right)^{2}}{2\sqrt{v_{z}^{2}k_{z}^{2}+v_{\perp}^{2}k_{\perp}^{2}}}+\frac{1}{2u_{c}}=0.
\end{align}
Then, the variation is
\begin{align}
\notag\delta\mathcal{F}_{\text{MF}}(\phi)=&\frac{\phi^{2}}{2u}-\frac{\phi^{2}}{2u_{c}}+\frac{1}{V}\sum_{\bm{k}}\left( E_{\bm{k}}(0)+\frac{\phi^{2}}{2}\frac{\left(\mathcal{F}(\theta_{\bm{k}})\right)^{2}}{\sqrt{v_{z}^{2}k_{z}^{2}+v_{\perp}^{2}k_{\perp}^{2}}}- E_{\bm{k}}(\phi) \right)-\frac{2T}{V}\sum_{\bm{k}}\ln\left(\frac{1+e^{-\beta E_{\bm{k}}(\phi)}}{1+e^{-\beta E_{\bm{k}}(0)}}\right)\\
\notag=&\frac{\phi^{2}}{2u}-\frac{\phi^{2}}{2u_{c}}+{2ck_{f}}|\phi|^{3}\int_{0}^{2\pi}\frac{\pi}{3}\left| \mathcal{F}(\theta_{k}) \right|^{3}+2T\sum_{\bm{k}}\frac{\beta\left(\mathcal{F}(\theta_{k})\right)^{2}}{e^{\beta E_{\bm{k}}(0)}+1}\frac{\phi^{2}}{2E_{\bm{k}}(0)}+\cdots\\
\notag=&\frac{\phi^{2}}{2u}-\frac{\phi^{2}}{2u_{c}}+{c'k_{f}}|\phi|^{3}+2c k_f \int d\theta_{k}\left(\mathcal{F}(\theta_{k})\right)^{2}\int_{k_{\perp},k_{z}}\frac{1}{e^{\beta E_{k}(0)}+1}\frac{\phi^{2}}{E_{k}(0)}+\cdots\\
\notag=&\frac{\phi^{2}}{2u}-\frac{\phi^{2}}{2u_{c}}+{c'k_{f}}|\phi|^{3}+\frac{2ck_{F}}{v_{\perp}v_{z}}T\int d\theta_{k}\left(\mathcal{F}(\theta_{k})\right)^{2}\int dk\;k\frac{1}{e^{ k}+1}\frac{\phi^{2}}{k}+\cdots\\
=&\left( \frac{u_{c}-u}{2u_{c}^{2}}+\tilde{c}T \right)\phi^{2}+c'k_{f}|\phi|^{3}+\cdots,
\end{align}
where 
\begin{align}
c'=&\frac{2 c\pi }{3}\int_{0}^{2\pi}d\theta_{k}\left|\mathcal{F}(\theta_{k})\right|^{3},\\
\tilde{c}=&\frac{2c\pi\ln2}{v_{\perp}v_{z}}\int_{0}^{2\pi}d\theta_{k}\left( \mathcal{F}(\theta_{k}) \right)^{2}, 
\end{align}
with $c = \frac{1}{(2\pi)^3}$. 

\section{Critical Theory}
We derive the critical theory $\mathcal{S}= \mathcal{S}_\psi + \mathcal{S}_\phi + \mathcal{S}_{\text{int}}$ appearing in the main text. We first start with the defintion of $\mathcal{S}_\psi$ which is 
\begin{align}
\mathcal{S}_\psi = \int_{\bm{k},\omega} \Psi^{\dagger}_{\bm{k},\omega}\Big(-i\omega + (v_z \delta k_z + \zeta \delta k_\perp \mu^z )\tau^z \mu^z + v_\perp \delta k_\perp \tau^x \Big)\Psi_{\bm{k}, \omega},  
\end{align}
where 
\begin{align}
\int_{\bm{k}, \omega} = \int \frac{d^3 k d\omega}{(2\pi)^4} \approx \int \frac{d \delta k_z}{2\pi} \int k_f \frac{d\delta k_\perp}{2\pi} \int d\theta_{\bm{k}} \int \frac{d\omega}{2\pi}, 
\label{FermionIntegral}
\end{align} 
i.e., the integral over the fermion momentum is assuming the kinematic structure of fermion near the nodal ring and incorporates only the fluctuation near the nodal ring.

\begin{figure}[h]
\begin{center}
\includegraphics[width=0.5\columnwidth]{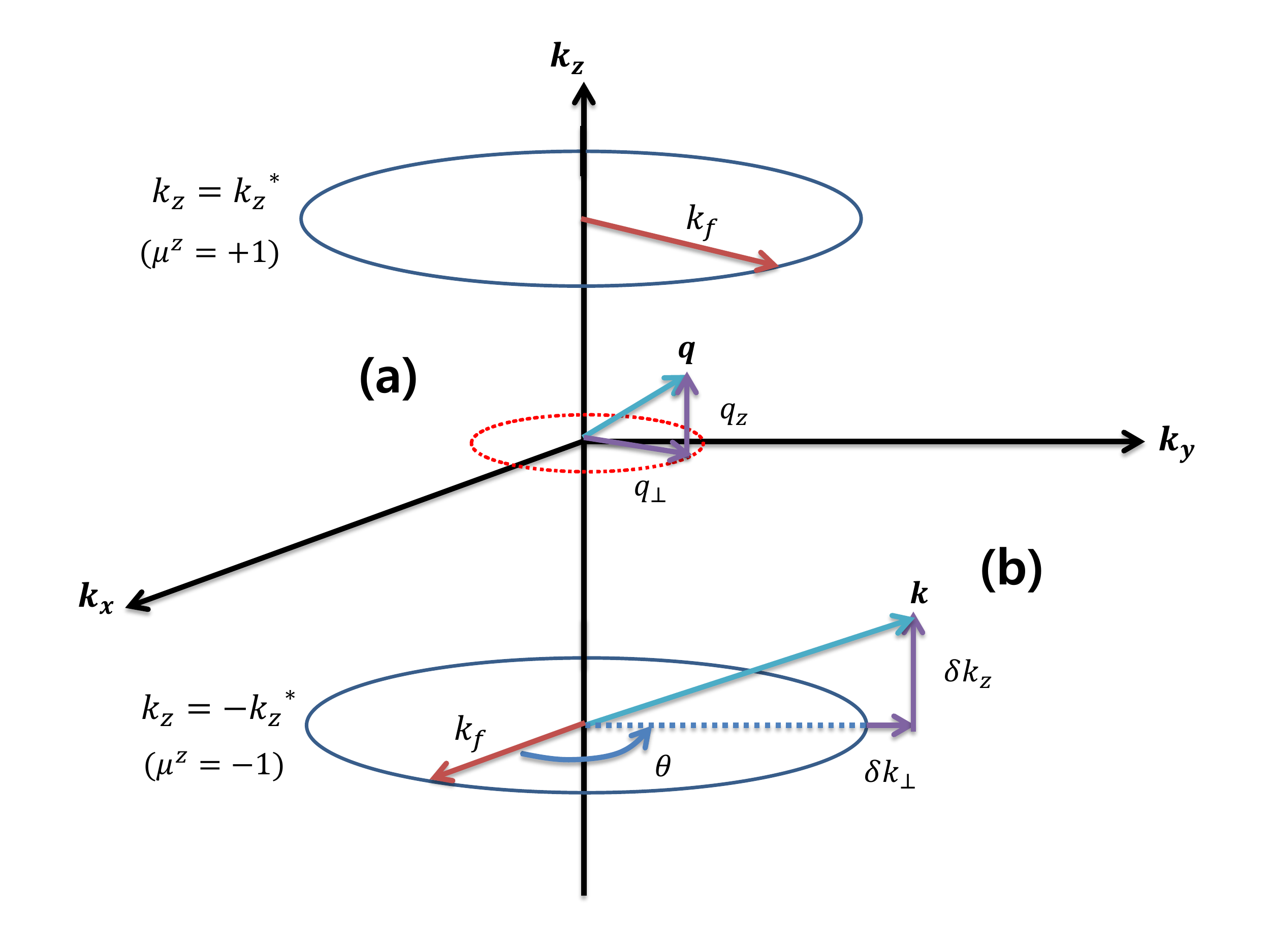}
\caption{
Line nodes and low-energy excitation modes. (a) is the range of the integrals for momentum of the order parameters. (b) is the range of the integral for momentum of the low-energy BdG fermions. Note that the fermion is fluctuating near the nodal line.  
 \label{Momentum}
}
\end{center}
\end{figure}

On the other hand, the interaction between the fermion and the boson $\mathcal{S}_{\text{int}}$ is 
\begin{align}
\mathcal{S}_{\text{int}} = g \int_{\bm{q}, \Omega}  \phi(\bm{q}, \Omega) \int_{\bm{k},\omega} \mathcal{F}(\bm{k};\bm{q}) \Psi^{\dagger}_{\bm{k}+\bm{q}, \omega+\Omega} \mathcal{M} \Psi_{\bm{k}, \omega},  
\end{align}
in which the integral over the bosonic momentum is defined as 
\begin{align}
\int_{\bm{q}, \Omega} = \int \frac{d^3 q d\omega}{(2\pi)^4},  
\end{align}
which is centered around the origin of momentum space. On the other hand, the integral over $\{\bm{k},\omega \}$ follows the same scheme as Eq.\eqref{FermionIntegral} because $\{\bm{k}, \omega\}$ are the momentum and frequency of the fermion. The difference in kinematics of the fluctuation of the bosons and low-energy BdG fermions is illustrated in Fig. \ref{Momentum}.

Throughout the main text and this supplemental material, we are interested in the limit $|\bm{q}| \ll k_f$ (with the size of nodal ring $k_f$) and hence use the approximation on the form factor $\mathcal{F}(\bm{k};\bm{q})\Big|_{|\bm{q}| \ll k_f}\approx \mathcal{F}(\bm{k};0) + O(\frac{|\bm{q}|}{k_f}) \to \mathcal{F}(\theta_{\bm{k}})$ upon projecting to the lowest-energy fermions near the nodal ring. Thus we finally end up with 
\begin{align}
\mathcal{S}_{\text{int}} \approx g \int_{\bm{q}, \Omega}  \phi(\bm{q}, \Omega) \int_{\bm{k},\omega} \mathcal{F}(\theta_{\bm{k}}) \Psi^{\dagger}_{\bm{k}+\bm{q}, \omega+\Omega} \mathcal{M} \Psi_{\bm{k}, \omega}, 
\end{align}
in the low-energy limit. As illustrated in the supplemental material \ref{Continuum}, the coupling between the order parameter and low-energy BdG Hamiltonian can be represented as \newline{}\newline
\textbf{1}.\textit{$A_{1}$ representation}: $\mathcal{F} (\theta) = 1$ and $\mathcal{M}=\tau^y$. \newline
\textbf{2}.\textit{$A_{2}$ representation}: $\mathcal{F} (\theta) = \sin(4\theta)$ and $\mathcal{M}=\tau^y$.\newline
\textbf{3}.\textit{$B_{1}$ representation}: $\mathcal{F} (\theta)= \cos(2\theta)$ and $\mathcal{M}=\tau^y$.\newline 
\textbf{4}.\textit{$B_{2}$ representation}: $\mathcal{F} (\theta)= \sin(2\theta)$ and $\mathcal{M}=\tau^y$.\newline
\textbf{5}.\textit{$E$ representations}: $\mathcal{F}_x(\theta) = \cos(\theta)$ or $\mathcal{F}_y(\theta) = \sin(\theta)$  with $\mathcal{M}=\mu^z \tau^y$. \newline{}\newline

On the other hand, the boson part $\mathcal{S}_\phi$ is 
\begin{align}
\mathcal{S}_\phi = \int_{\bm{k},\omega} \frac{1}{2}(|\bm{k}|^2 + \omega^2 + r) |\phi_{\bm{k}, \omega}|^2. 
\end{align}

Because of $[\mathcal{M}, \mu^z]=0, \mathcal{M} = \tau^y, \tau^y \mu^z$, it is useful to perform the decomposition $\Psi^{\dagger}_{\bm{k},\omega} = \Big[ \chi^{\dagger}_{\bm{k},\omega, \uparrow}, \chi^{\dagger}_{\bm{k},\omega, \downarrow} \Big]$ such that $\mu^z \chi{\dagger}_{\uparrow/\downarrow} = \pm \chi^{\dagger}_{\uparrow/\downarrow}$, i.e., $\chi_\uparrow$-fermion ($\chi_\downarrow$-fermion) represents the fermions near the upper nodal ring $k_z \approx k_z^{*}$ (near the lower nodal ring $k_z \approx- k_z^{*}$). 

With this in hand, 
\begin{align}
\mathcal{S}_\psi &= \int_{\bm{k}, \omega} \chi^{\dagger}_{\bm{k},\omega,\uparrow} \Big[ -i\omega + (v_z \delta k_z + \zeta \delta k_\perp)\tau^z + v_\perp \delta k_\perp \tau^x \Big]  \chi_{\bm{k},\omega,\uparrow} \nonumber\\ 
&+ \int_{\bm{k}, \omega} \chi^{\dagger}_{\bm{k},\omega,\downarrow} \Big[ -i\omega + (v_z \delta k_z - \zeta \delta k_\perp)(-\tau^z) + v_\perp \delta k_\perp \tau^x \Big]  \chi_{\bm{k},\omega,\downarrow}, 
\end{align}
and 
\begin{align}
\mathcal{S}_{\text{int}} = g \int_{\bm{q},\Omega} \phi_{\bm{q},\Omega} \int_{\bm{k}, \omega} \mathcal{F}(\theta_{\bm{k}})  \Big[ \chi^{\dagger}_{\bm{k}+\bm{q},\omega+\Omega, \uparrow}\tau^y \chi_{\bm{k},\omega, \uparrow} + \sigma \chi^{\dagger}_{\bm{k}+\bm{q},\omega+\Omega, \downarrow}\tau^y \chi_{\bm{k},\omega, \downarrow} \Big],
\end{align}
in which $\sigma = +1$ for $\mathcal{M} = \tau^y$ and $\sigma = -1$ for $\mathcal{M} = \tau^y \mu^z$. We see that $\chi_\uparrow$-fermion is decoupled from $\chi_\downarrow$-fermion within the effective theory. 

We now perform the shift of the momentum for the fermions as following 
\begin{align}
&v_z \delta k_z' = v_z \delta k_z + \zeta \delta k_\perp, ~ v_\perp \delta k_\perp' = v_\perp \delta k_\perp, ~ \theta_{\bm{k}'} = \theta_{\bm{k}} \text{ for  $\chi_\uparrow$-fermion},\nonumber\\ 
&v_z \delta k_z' = v_z \delta k_z - \zeta \delta k_\perp, ~ v_\perp \delta k_\perp' = v_\perp \delta k_\perp,~ \theta_{\bm{k}'} = \theta_{\bm{k}}  \text{ for  $\chi_\downarrow$-fermion}. 
\label{Transformation}
\end{align}
This transformation will lead us to the following critical theory $\mathcal{S} = \mathcal{S}_\psi + \mathcal{S}_{\text{int}} + \mathcal{S}_\phi$ 
\begin{align}
& \mathcal{S}_\psi = \int_{\bm{k},\omega} \Psi^{\dagger}_{\bm{k},\omega}\Big(-i\omega + v_z \delta k_z \tau^z \mu^z + v_\perp \delta k_\perp \tau^x \Big)\Psi_{\bm{k}, \omega}, \nonumber\\ 
& \mathcal{S}_{\text{int}} = g \int_{\bm{q}, \Omega}  \phi(\bm{q}, \Omega) \int_{\bm{k},\omega} \mathcal{F}(\theta_{\bm{k}}) \Psi^{\dagger}_{\bm{k}+\bm{q}, \omega+\Omega} \mathcal{M} \Psi_{\bm{k}, \omega},  \nonumber\\ 
& \mathcal{S}_\phi = \int_{\bm{k},\omega} \frac{1}{2}(|\bm{k}|^2 + \omega^2 + r) |\phi_{\bm{k}, \omega}|^2, 
\label{CriticalTheory1}
\end{align} 
in which we have relabelled $\bm{k}' \to \bm{k}$ and recombined $\Big[\chi^{\dagger}_{\bm{k},\omega,\uparrow},\chi^{\dagger}_{\bm{k},\omega, \downarrow} \Big] \to \Psi^{\dagger}_{\bm{k},\omega}$. The only effect of the transformation Eq.\eqref{Transformation} is $(v_z \delta k_z + \zeta \delta k_\perp \mu^z )\tau^z \mu^z \to v_z \delta k_z \tau^z \mu^z$ in $\mathcal{S}_\psi$. Furthermore, the critical theory takes the same form irrespective of $\mathcal{M} = \tau^y$ or $\mathcal{M}=\tau^y \mu^z$. Hence, we first restrict ourselves to the case $\mathcal{M} = \tau^y$, which corresponds to the one-dimensional representations of the symmetry group, and then we will extend to the two-dimensional representation where $\mathcal{M} = \tau^y \mu^z$ appears.  

Furthermore, because the two nodal-ring fermions generate the same corrections to the boson and fermion self-energies, hereafter we take only the upper nodal-ring fermions, i.e., of $\mu^z = +1$, to discuss the critical theory $\mathcal{S} = \mathcal{S}_\psi + \mathcal{S}_{\text{int}} + \mathcal{S}_\phi$ 
\begin{align}
& \mathcal{S}_\psi = \int_{\bm{k},\omega} \Psi^{\dagger}_{\bm{k},\omega}\Big(-i\omega + v_z \delta k_z \tau^z  + v_\perp \delta k_\perp \tau^x \Big)\Psi_{\bm{k}, \omega}, \nonumber\\ 
& \mathcal{S}_{\text{int}} = g \int_{\bm{q}, \Omega}  \phi(\bm{q}, \Omega) \int_{\bm{k},\omega} \mathcal{F}(\theta_{\bm{k}}) \Psi^{\dagger}_{\bm{k}+\bm{q}, \omega+\Omega} \mathcal{M} \Psi_{\bm{k}, \omega},  \nonumber\\ 
& \mathcal{S}_\phi = \int_{\bm{k},\omega} \frac{1}{2}(|\bm{k}|^2 + \omega^2 + r) |\phi_{\bm{k}, \omega}|^2, 
\label{CriticalTheory}
\end{align} 
while keeping in mind that integrating out the fermion propagator comes with the factor of $2$, i.e., there are effectively two flavors of the fermion. 

\begin{figure}[h]
\begin{center}
\includegraphics[width=0.5\columnwidth]{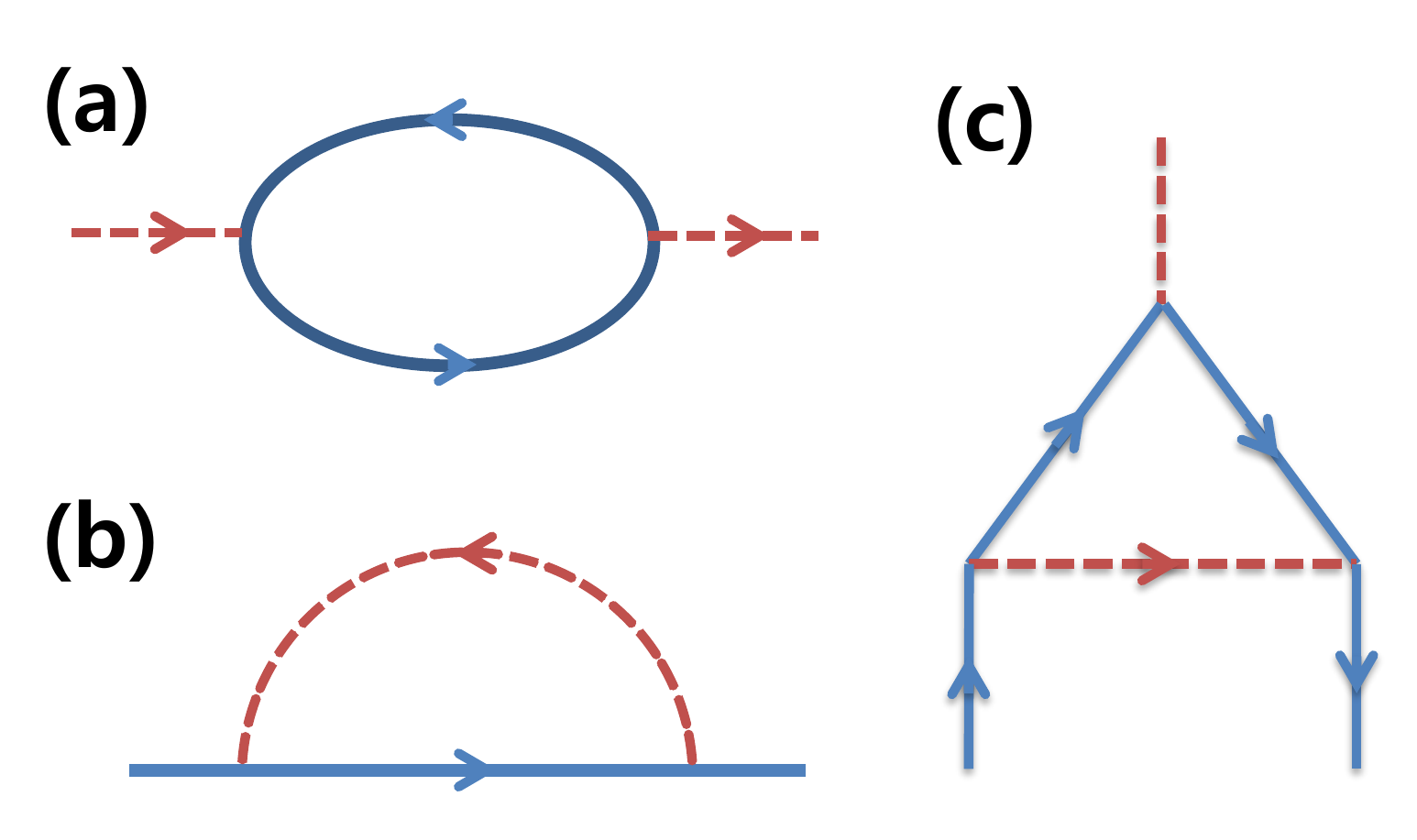}
\caption{
Feynman Diagrams. The dotted line represents the order parameter propagator and the solid line represents the fermion propagator. (a) Boson self-energy. (b) Fermion self-energy. (c) Vertex correction. 
 \label{Diagrams}
}
\end{center}
\end{figure}

To perform the renormalization analysis, below we will calculate the three Feynman diagrams, boson self-energy, fermion self-energy and the vertex corrections as in Fig. \ref{Diagrams}. 

\subsection{Boson Self-Energy} 
From the critical theory Eq.\eqref{CriticalTheory}, we compute the boson self-energy in the standard large-$N_f$ limit where $2 \times N_f$ is the number of the flavors of the fermions coupled to the boson (the factor of $2$ comes from the fact that there are two nodal rings per each flavor).   
\begin{align}
\Sigma_b(\bm{q}, \Omega) = g^2 \int_{\bm{k},\omega} \mathcal{F}(\theta_{\bm{k}})  \mathcal{F}(\theta_{\bm{k}+\bm{q}}) \text{Tr}\Big[ \tau^y  G_{f,0} (\omega, \bm{k}) \tau^y G_{f,0}(\omega+\Omega, \bm{k}+ \bm{q}) \Big], 
\end{align}
in which $G_{f,0}^{-1}(\bm{k},\omega) = -i\omega + v_z \delta k_z \tau^z + v_\perp \delta k_\perp \tau^x$ is used (Here the `Tr$[\cdot]$' is acting on the $N_f$-flavor space, too). Here we align $\bm{q} = (q_\perp, 0, q_z)$, i.e., $\bm{q}_\perp$ is aligned along $\hat{x}$-axis. However, due to the rotational symmetry, the final result of $\Sigma_b (\bm{q}, \Omega)$ will apply any $\bm{q}$. Now Using 
\begin{align}
\mathcal{F}(\theta_{\bm{k} + \bm{q}}) \approx \mathcal{F}(\theta_{\bm{k}}) + O(\frac{|\bm{q}|}{k_f}), \text{ where }~ \frac{|\bm{q}|}{k_f} \ll 1, 
\end{align}
we find 
\begin{align}
\Sigma_b(\bm{q}, \Omega) = g^2 \int_{\bm{k},\omega} \Big(\mathcal{F}(\theta_{\bm{k}})\Big)^2  \text{Tr}\Big[ \tau^y  G_{f,0} (\omega, \bm{k}) \tau^y G_{f,0}(\omega+\Omega, \bm{k}+ \bm{q}) \Big]. 
\end{align}
Plugging the integration measure Eq.\eqref{FermionIntegral} explicitly, we further find 
\begin{align}
\Sigma_b(\bm{q}, \Omega) &= g^2 k_f\int \frac{d\theta_{\bm{k}}}{2\pi} \Big(\mathcal{F}(\theta_{\bm{k}})\Big)^2  \int \frac{d \delta k_z d \delta k_\perp d\omega}{(2\pi)^3} \text{Tr}\Big[ \tau^y  G_{f,0} (\omega, \bm{k}) \tau^y G_{f,0}(\omega+\Omega, \bm{k}+ \bm{q}) \Big] \nonumber\\
&=g^2 k_f \int \frac{d\theta_{\bm{k}}}{2\pi} \Big(\mathcal{F}(\theta_{\bm{k}})\Big)^2 J[\bm{q}, \Omega; \theta_{\bm{k}}]. 
\end{align}
Here $J[\bm{q}, \Omega; \theta_{\bm{k}}]$ is the integral 
\begin{align}
J[\bm{q}, \Omega; \theta_{\bm{k}}] = \int \frac{d \delta k_z d \delta k_\perp d\omega}{(2\pi)^3} \text{Tr}\Big[ \tau^y  G_{f,0} (\omega, \bm{k}) \tau^y G_{f,0}(\omega+\Omega, \bm{k}+ \bm{q}) \Big].
\end{align}
We evaluate the integral $J[\bm{q}, \Omega; \theta_{\bm{k}}]$ first. 
\begin{align}
J[\bm{q}, \Omega; \theta_{\bm{k}}]&=\int \frac{d \delta k_z d \delta k_\perp d\omega}{(2\pi)^3} \text{Tr}\Big[ \tau^y  G_{f,0} (\omega, \bm{k}) \tau^y G_{f,0}(\omega+\Omega, \bm{k}+ \bm{q}) \Big] \nonumber\\ 
&= \int \frac{d \delta k_z d \delta k_\perp d\omega}{(2\pi)^3} \text{Tr} \Big[\frac{1}{-i\omega- v_z k_z \tau^x - v_\perp \delta k_\perp \tau^z} \frac{1}{-i(\omega+\Omega) + v_z (q_z+ \delta k_z) \tau^x + v_\perp (\delta k_\perp + q_\perp \cos(\theta_{\bm{k}})) \tau^z}\Big].
\end{align}
Now we reexpress the integral in terms of the variables $\bm{Q} = (\omega_n, v_z \delta k_z , v_\perp \delta k_\perp), ~ \bm{P} = (\Omega_n, v_z q_z, v_\perp q_\perp \cos(\theta_{\bm{k}}))$  to find 
\begin{align}
J[\bm{q}, \Omega; \theta_{\bm{k}}] = -\frac{2 N_f}{v_\perp v_z} \int \frac{dQ^3}{(2\pi)^3} \Big[\frac{\bm{Q}\cdot (\bm{Q}+\bm{P})}{\bm{Q}^2 (\bm{Q}+\bm{P})^2}\Big] = \frac{2 N_f}{16 v_\perp v_z} |\bm{P}| = \frac{ N_f}{8 v_\perp v_z}  [\Omega^2 + v^2_z q^2_z + v_\perp^2 q_\perp^2 \cos^2 (\theta_{\bm{k}}) ]^{1/2} ,
\end{align}
where the integral over $\bm{Q}$ can be performed analytically by the conventional dimensional regularization. This dimensional regularization automatically subtract out the divergent contribution to the boson self-energy. Hence we finally find 
\begin{align}
\Sigma_b(\bm{q}, \Omega) = \frac{g^2 k_f N_f}{16\pi v_\perp v_z} \int^{2\pi}_{0} d\theta \Big(\mathcal{F}(\theta)\Big)^2 \Big[\Omega^2 + v_z^2 q_z^2 + v_\perp^2 q_\perp^2 \cos^2 (\theta)\Big]^{1/2}. 
\label{BosonSelfE}
\end{align}
At this stage, it is worth to note the followings. First of all, there is no divergence in performing the integral over $\theta$ in boson self-energy Eq.\eqref{BosonSelfE} because $\mathcal{F}^2(\theta)$ and $\Big[\Omega^2 + v_z^2 q_z^2 + v_\perp^2 q_\perp^2 \cos^2 (\theta)\Big]^{1/2}$ are regular and bounded in $\theta \in [0, 2\pi]$ (rememeber that $\mathcal{F}(\theta) \sim \cos(n\theta), n \in \mathbb{Z}$ or $\sim \sin(n \theta), n \in \mathbb{Z}^{+}$).  Secondly, it is apparent that 
\begin{align}
|\Sigma_b(\bm{q}, \Omega)| \sim \Big[\Omega^2 + v_z^2 q_z^2 + v_\perp^2 q_\perp^2 \cos^2 (\theta)\Big]^{1/2} \gg g_{b,0}^{-1}(\bm{q}, \Omega) \sim |\bm{q}|^2 + \Omega^2, 
\end{align}
at the low-energy limit and long-distance limit, i.e., $\Omega \to 0$ and $\bm{q} \to \bm{0}$, at the critical point. Thus, being interested in the low-energy limit, we finally have 
\begin{align}
G_{b}^{-1}(\bm{q}, \Omega) = G_{b,0}^{-1}(\bm{q}, \Omega) + \Sigma_b (\bm{q}, \Omega) \approx  \Sigma_b (\bm{q}, \Omega), 
\end{align}
and hence we will use $G_b (\bm{q}, \Omega) \approx \frac{1}{\Sigma_b(\bm{q},\Omega)} \sim O(\tfrac{1}{N_f})$ as expected.  

The explicit forms for the boson self-energy for each representation are following. 
\begin{align}
\Sigma_{b}(\Omega,\mathbf{q})=&\frac{g^{2}N_{f}k_{f}}{16\pi v_{\perp}v_{z}}\int_{0}^{2\pi}d\theta\;\left(\mathcal{F}(\theta)\right)^{2}\sqrt{\Omega^{2}+v_{z}^{2}q_{z}^{2}+v_{\perp}^{2}q_{\perp}^{2}\cos^{2}\theta}\notag\\
=&\frac{g^{2}N_{f}k_{f}}{16\pi v_{\perp}v_{z}}\int_{0}^{2\pi}d\theta\;\left(\mathcal{F}(\theta)\right)^{2}\sqrt{\Omega^{2}+v_{z}^{2}q_{z}^{2}+v_{\perp}^{2}q_{\perp}^{2}-v_{\perp}^{2}q_{\perp}^{2}\sin^{2}\theta}\notag\\
=&\frac{g^{2}N_{f}k_{f}}{16\pi v_{\perp}v_{z}}\sqrt{\Omega^{2}+v_{z}^{2}q_{z}^{2}+v_{\perp}^{2}q_{\perp}^{2}}\int_{0}^{2\pi}d\theta\;\left(\mathcal{F}(\theta)\right)^{2}\sqrt{1-\frac{v_{\perp}^{2}q_{\perp}^{2}}{\Omega^{2}+v_{z}^{2}q_{z}^{2}+v_{\perp}^{2}q_{\perp}^{2}}\sin^{2}\theta}\notag\\
=&\frac{g^{2}N_{f}k_{f}}{16\pi v_{\perp}v_{z}}\sqrt{\Omega^{2}+v_{z}^{2}q_{z}^{2}+v_{\perp}^{2}q_{\perp}^{2}}\int_{0}^{2\pi}d\theta\;\left(\mathcal{F}(\theta)\right)^{2}\sqrt{1-\rho(\Omega,\mathbf{q})\sin^{2}\theta}\notag\\
=&k_{f}N_{f}\sqrt{\Omega^{2}+v_{z}^{2}q_{z}^{2}+v_{\perp}^{2}q_{\perp}^{2}}\mathcal{R}(\rho(\Omega,\mathbf{q})),
\end{align}
where  $\rho(\Omega,\mathbf{q})=1/\left(1+\frac{\Omega^{2}+v_{z}^{2}q_{z}^{2}}{v_{\perp}^{2}q_{\perp}^{2}}\right)$, $0\leq \rho\leq 1$, and
\begin{align}
\mathcal{R}(x)=\mathcal{C}\int_{0}^{\pi/2}d\theta\left(\mathcal{F}(\theta)\right)^{2}\sqrt{1-x\sin^{2}\theta},
\end{align}
with $\mathcal{C}=\frac{g^{2}}{4\pi v_{\perp}v_{z}}$. For each representation, $\mathcal{R}(x)$ is
\begin{align}
\mathcal{R}_{A_{1g}}(x)=&\mathcal{C}El_{2}(x),\\
\mathcal{R}_{A_{2g}}(x)=&\frac{16\mathcal{C}}{315x^{4}}((64-128x+78x^{2}-14x^{3}+10x^{4})El_{2}(x)+(-64+160x-138x^{2}+47x^{3}-5x^{4})El_{1}(x)),\\
\mathcal{R}_{B_{1g}}(x)=&\frac{\mathcal{C}}{15x^{2}}((-8+8x+7x^{2})El_{2}(x)+(8-12x+4x^{2})El_{1}(x)),\\
\mathcal{R}_{B_{2g}}(x)=&\frac{4\mathcal{C}}{15x^{2}}((2-2x+2x^{2})El_{2}(x)-(2-3x+x^{2})El_{1}(x)),\\
\mathcal{R}_{E}(x)=&\frac{\mathcal{C}}{3x}((1+x)El_{2}(x)+(-1+x)El_{1}(x)),\\
\mathcal{R}_{E'}(x)=&\frac{\mathcal{C}}{3x}((-1+2x)El_{2}(x)-(-1+x)El_{1}(x)),
\end{align}
where $El_{1}(x)$ and $El_{2}(x)$ are the complete elliptic integral of first and second kinds, respectively,
\begin{align}
El_{1}(x)=&\int_{0}^{\pi/2}\frac{d\theta}{\sqrt{1-x\sin^{2}\theta}},\\
El_{2}(x)=&\int_{0}^{\pi/2}\sqrt{1-x\sin^{2}\theta}d\theta.
\end{align}
\begin{figure}[h]
\begin{center}
\includegraphics[width=0.5\textwidth]{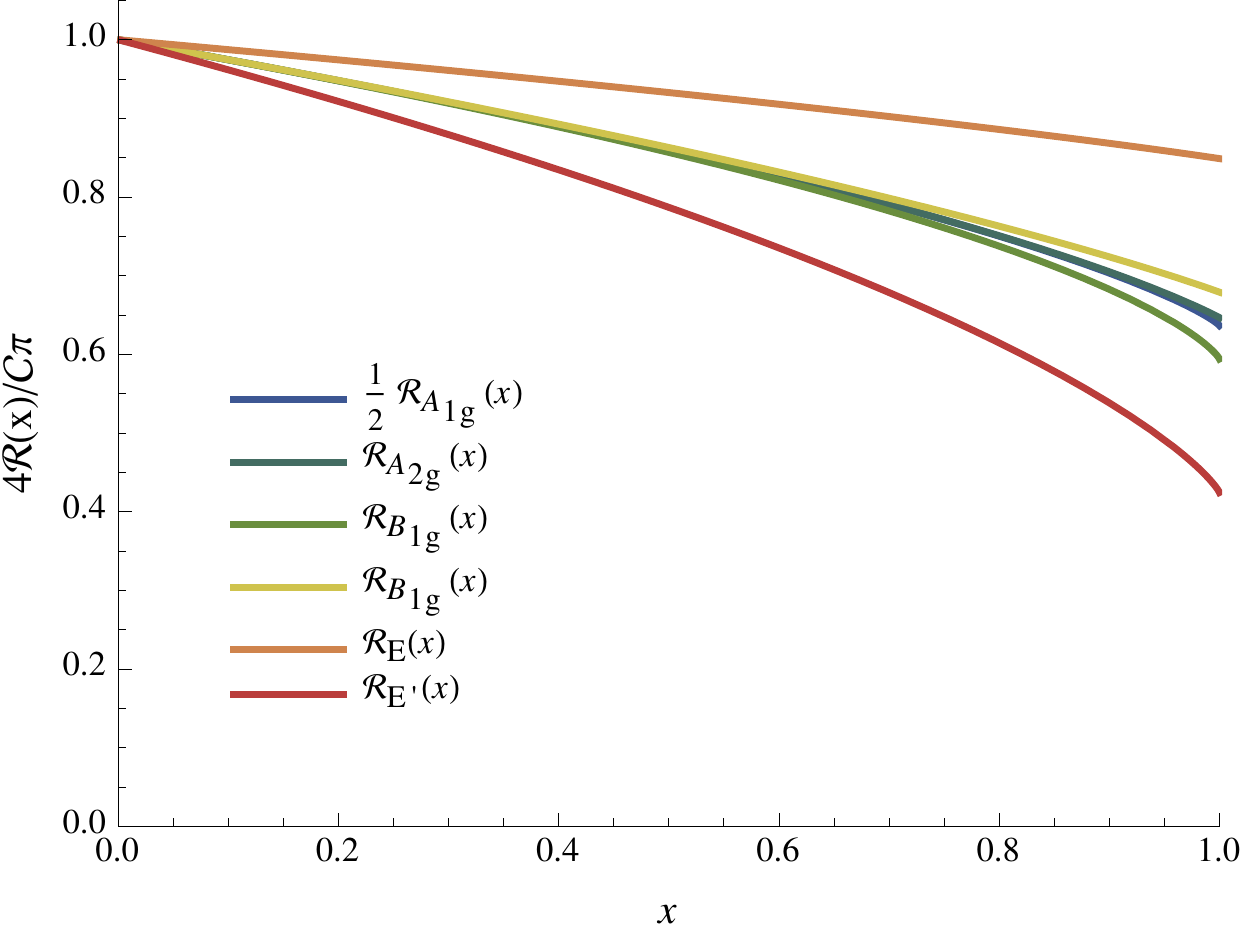}
\caption{$\mathcal{R}$ function for each representations.
 \label{smFig1}
}
\end{center}
\end{figure}

\subsubsection*{Boson Self-Energy with Non-linearized Hamiltonian}
From the linearized fermion dispersion near the line node, we have found, via analytic calculation, that the boson self-energy has the following ``schematic" form 
\begin{align}
|\Sigma_b (\bm{q}, \Omega)| \sim |\Omega| + v_z|q_z| + v_\perp |q_\perp|. \nonumber 
\end{align}
One may worry that this linear dependences of the boson self-energy on $\{\Omega, q_\perp, q_z\}$ are an artifact of the linearized fermion dispersion. Motivated by this, we now confirm that the behavior, i.e., the linear dependences of the boson self-energy on $\{\Omega, q_\perp, q_z\}$, is not an artifact and rather a robust feature of the dynamics of the order parameter coupled to a nodal line SC.  

\begin{figure}[h]
\begin{center}
\includegraphics[width=0.5\textwidth]{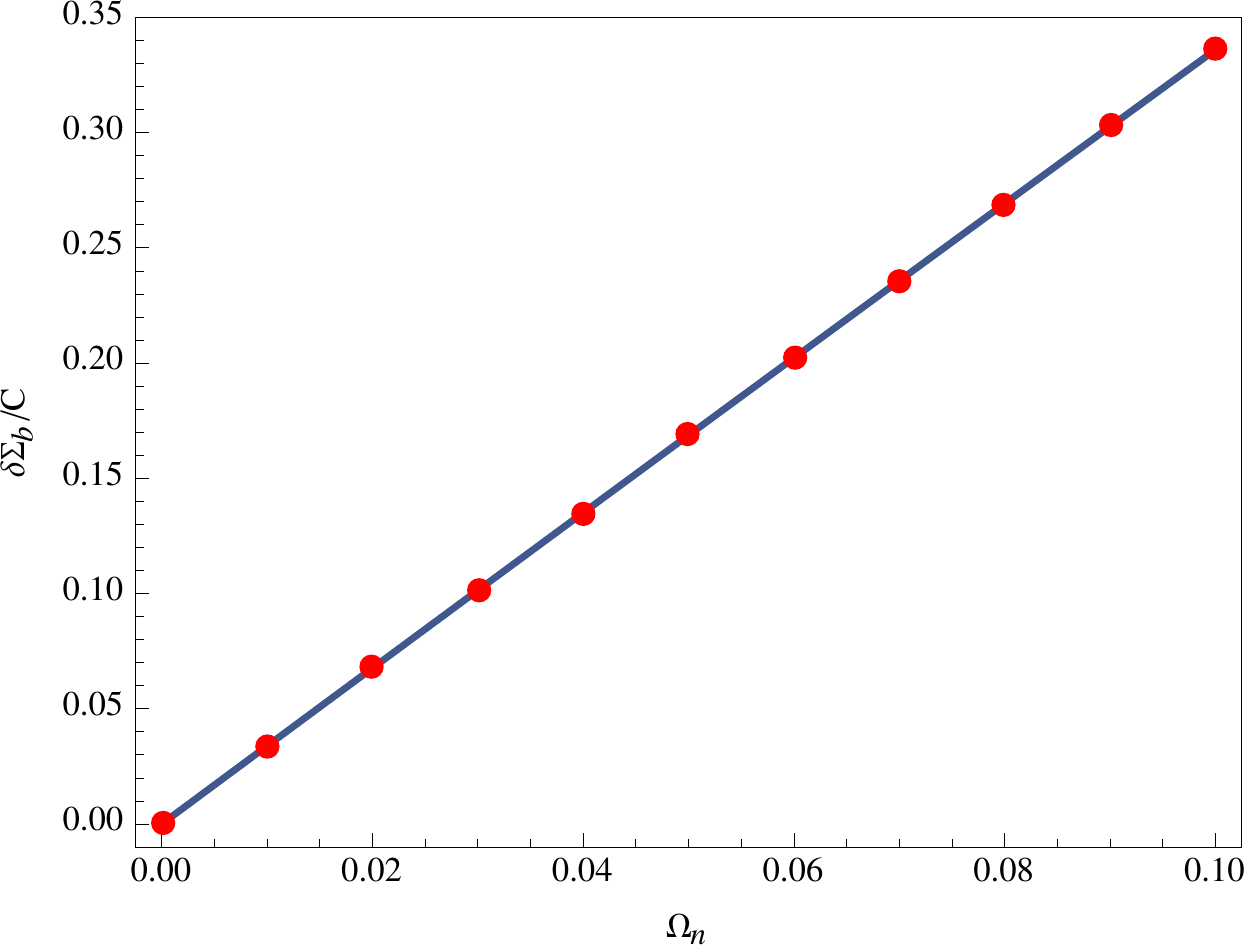}
\caption{Numerical Calculation of Boson Self-energy $\Sigma_b(\bm{0}, \Omega)$ normalized by $C=\frac{g^{2}}{(2\pi)^{3}}\frac{2\pi m}{v_{z}}$. Here we set $k_{F} = 1$, $m=\frac{1}{2}$ and $v_{z}=1$ for the numerical calculation. This clearly demonstrates $ \Sigma_b(\bm{0}, \Omega) \sim |\Omega|$. 
 \label{Om}
}
\end{center}
\end{figure}

To show this, we calculate the dependence of the boson self energy on $\{\Omega, q_\perp, q_z\}$ numerically, with the following Hamiltonian, which we have not linearized along $k_\perp$
\begin{align*}
\mathcal{H}_{0}=\frac{k_{x}^{2}+k_{y}^{2}-k_{F}^{2}}{2m}\tau^{x}+v_{z}k_{z}\tau^{z}\equiv\epsilon_{n}(\mathbf{k})\tau^{n},\quad n=x,z.
\end{align*}
The energy spectrum is $E(\mathbf{k})=\pm\sqrt{\frac{(k_{x}^{2}+k_{y}^{2}-k_{F}^{2})^{2}}{4m^{2}}+v_{z}^{2}k_{z}^{2}}$ and the fermion propagator is
\begin{align*}
G(\mathbf{k},i\omega_{n})=\frac{1}{-i\omega_{n}+\mathcal{H}_{0}(\mathbf{k})}=\frac{1}{-i\omega_{n}+E_{\alpha}(\mathbf{k})}\text{P}_{\alpha}(\mathbf{k}),
\end{align*}
where $\text{P}_{\alpha}(\mathbf{k})=\frac{1}{2}(1+\alpha \frac{\mathcal{H}_{0}(\mathbf{k})}{|E(\mathbf{k})|})$ with $\alpha=+$ or $-$. For simplicity, here we only conisder the self-energy crrection for the $A_{1}$ representation, $\mathcal{F}(\theta)=1$ (however, it is straightforward to generalize to the other representations). The boson self-energy is
\begin{align*}
\Sigma_{b}(\Omega_{n},\mathbf{q})=g^{2}\int \frac{d^{3}k}{(2\pi)^{3}}\text{Tr}(\text{P}_{\alpha}(\mathbf{k}+\mathbf{q}/2)\tau^{y}\text{P}_{\gamma}(\mathbf{k}-\mathbf{q}/2)\tau^{y})\frac{n_{F}(\alpha E(\mathbf{k}+\mathbf{q}/2))-n_{F}(\gamma E(\mathbf{k}-\mathbf{q}/2))}{-i\Omega_{n}+\alpha E(\mathbf{k}+\mathbf{q}/2)-\gamma E(\mathbf{k}-\mathbf{q}/2)},
\end{align*}
where $\alpha,\gamma=+$ or $-$, $n_{F}=1/(e^{\beta x}+1)$, and $\beta=1/k_{B}T$. This expression has the divergence which should be considered as the renormalization of the mass term of the order parameter, and we will subtract the divergent part out and extract the finite part to extract the dynamics of the order parameter at the quantum criticality.  
\newline{}\newline
\textbf{a. Frequency Dependence}: We start to evaluate the dependence of the boson self-energy on the frequency $\Omega$ to confirm 
\begin{align}
\Sigma_b (0, \Omega) \sim |\Omega|. \nonumber
\end{align}
We first evaluate the boson self-energy with $\bm{q} = 0$ 
\begin{align}
\notag\Sigma_{b}(\Omega_{n},0)=&\frac{g^{2}}{(2\pi)^{3}}\int d^{3}k\text{Tr}\left[\text{P}_{\alpha}(\mathbf{k})\tau^{y}\text{P}_{\gamma}(\mathbf{k})\tau^{y}\right]\frac{n_{F}(\alpha E(\mathbf{k}) )-n_{F}(\gamma E(\mathbf{k}))}{-i\Omega_{n}+\alpha E(\mathbf{k})-\gamma E(\mathbf{k})},\\
=&\frac{g^{2}}{(2\pi)^{3}}\int d^{3}k  (1-2n_{F}( E(\mathbf{k}) )) \left( -\frac{4E(\mathbf{k})}{\Omega_{n}^{2}+4E(\mathbf{k})^{2}}\right),
\end{align}
where 
\begin{equation}
\begin{aligned}
\text{Tr}[\text{P}_{\pm}(\mathbf{k})\tau^{y}\text{P}_{\pm}(\mathbf{k})\tau^{y}]=&0,\\
\text{Tr}[\text{P}_{\pm}(\mathbf{k})\tau^{y}\text{P}_{\mp}(\mathbf{k})\tau^{y}]=&1.
\end{aligned}
\end{equation}
At the zero temperature limit, we have $\beta\rightarrow\infty$,
\begin{align}
\Sigma_{b}(\Omega_{n},0)=&-\frac{g^{2}}{(2\pi)^{3}}\int d^{3}k \frac{4E(\mathbf{k})}{\Omega_{n}^{2}+4E(\mathbf{k})^{2}}. 
\end{align}
As prescribed above, we now subtract the divergent part of the integral to extract the finite contribution which describes the dynamics of the order parameter. Then,  
\begin{align}
\notag\delta\Sigma_{b}(\Omega_{n},0)=&-\frac{g^{2}}{(2\pi)^{3}}\int d^{3}k \left(\frac{4E(\mathbf{k})}{\Omega_{n}^{2}+4E(\mathbf{k})^{2}}-\frac{1}{E(\mathbf{k})}\right),\\
\notag=&\frac{g^{2}}{(2\pi)^{3}}\int d^{3}k \frac{1}{E(\mathbf{k})}\frac{\Omega_{n}^{2}}{\Omega_{n}^{2}+4E(\mathbf{k})^{2}},\\
=&\frac{g^{2}}{(2\pi)^{3}}\frac{2\pi m}{v_{z}}\int_{0}^{\infty}d\epsilon_{\perp}\int_{-\infty}^{\infty}d\epsilon_{z} \frac{1}{\sqrt{(\epsilon_{\perp}-\epsilon_{F})^{2}+\epsilon_{z}^{2}}}\frac{\Omega_{n}^{2}}{\Omega_{n}^{2}+4((\epsilon_{\perp}-\epsilon_{F})^{2}+\epsilon_{z}^{2})},
\end{align}
where $\epsilon_{\perp}=(k_{x}^{2}+k_{y}^{2})/2m$, $\epsilon_{F}=k_{F}^{2}/2m$ and $\epsilon_{z}=v_{z}k_{z}$. Shifting $\epsilon_\perp$ by $\epsilon_F$, we find
\begin{align}
\delta\Sigma_{b}(\Omega_{n},0)=&\frac{g^{2}}{(2\pi)^{3}}\frac{2\pi m}{v_{z}}\int_{-\epsilon_{F}}^{\infty}d\epsilon_{\perp}\int_{-\infty}^{\infty}d\epsilon_{z} \frac{1}{\sqrt{\epsilon_{\perp}^{2}+\epsilon_{z}^{2}}}\frac{\Omega_{n}^{2}}{\Omega_{n}^{2}+4(\epsilon_{\perp}^{2}+\epsilon_{z}^{2})}. 
\end{align}
We next perform the scaling of the variables, $\epsilon_{\perp}=|\Omega_{n}|x$ and $\epsilon_{z}=|\Omega_{n}|z$ to find 
\begin{align}
\delta\Sigma_{b}(\Omega_{n},0)=&\frac{g^{2}}{(2\pi)^{3}}\frac{2\pi m}{v_{z}}|\Omega_{n}|\int_{-\epsilon_{F}/|\Omega_{n}|}^{\infty}dx\int_{-\infty}^{\infty}dz \frac{1}{\sqrt{x^{2}+z^{2}}}\frac{1}{1+4(x^{2}+z^{2})}.
\end{align}
For the low-frequency limit, i.e., $\epsilon_{F}\gg |\Omega_{n}|$, we can approximate $-\frac{\epsilon_{F}}{|\Omega_{n}|}\to -\infty$ to find 
\begin{align}
\notag\delta\Sigma_{b}(\Omega_{n},0)=&\frac{g^{2}}{(2\pi)^{3}}\frac{2\pi m}{v_{z}}|\Omega_{n}|\int_{-\infty}^{\infty}dx\int_{-\infty}^{\infty}dz \frac{1}{\sqrt{x^{2}+z^{2}}}\frac{1}{1+4(x^{2}+z^{2})},\\
=&\frac{g^{2}}{(2\pi)^{3}}\frac{2\pi m}{v_{z}}C_{\Omega}|\Omega_{n}|,
\end{align}
where $C_{\Omega}$ is
\begin{align}
C_{\Omega}=\int_{-\infty}^{\infty}dx\int_{-\infty}^{\infty}dz \frac{1}{\sqrt{x^{2}+z^{2}}}\frac{1}{1+4(x^{2}+z^{2})}=4.9348. 
\end{align}
Hence we clearly see $\Sigma_b (0, \Omega) \sim |\Omega|$. We can also perform a numerical integral to confirm this behavior in Fig. \ref{Om}.  

\textbf{b. $q_{z}$ Dependence}: Next we calculate the dependence of the boson self-energy on $q_z$. To see this, we set  $\bm{q}=(0,0,q_{z})$ with $\Omega =0$. We expect to see 
\begin{align}
\Sigma_b (\bm{q}, 0) \sim |q_z|. \nonumber
\end{align}

\begin{figure}[h]
\begin{center}
\includegraphics[width=0.5\textwidth]{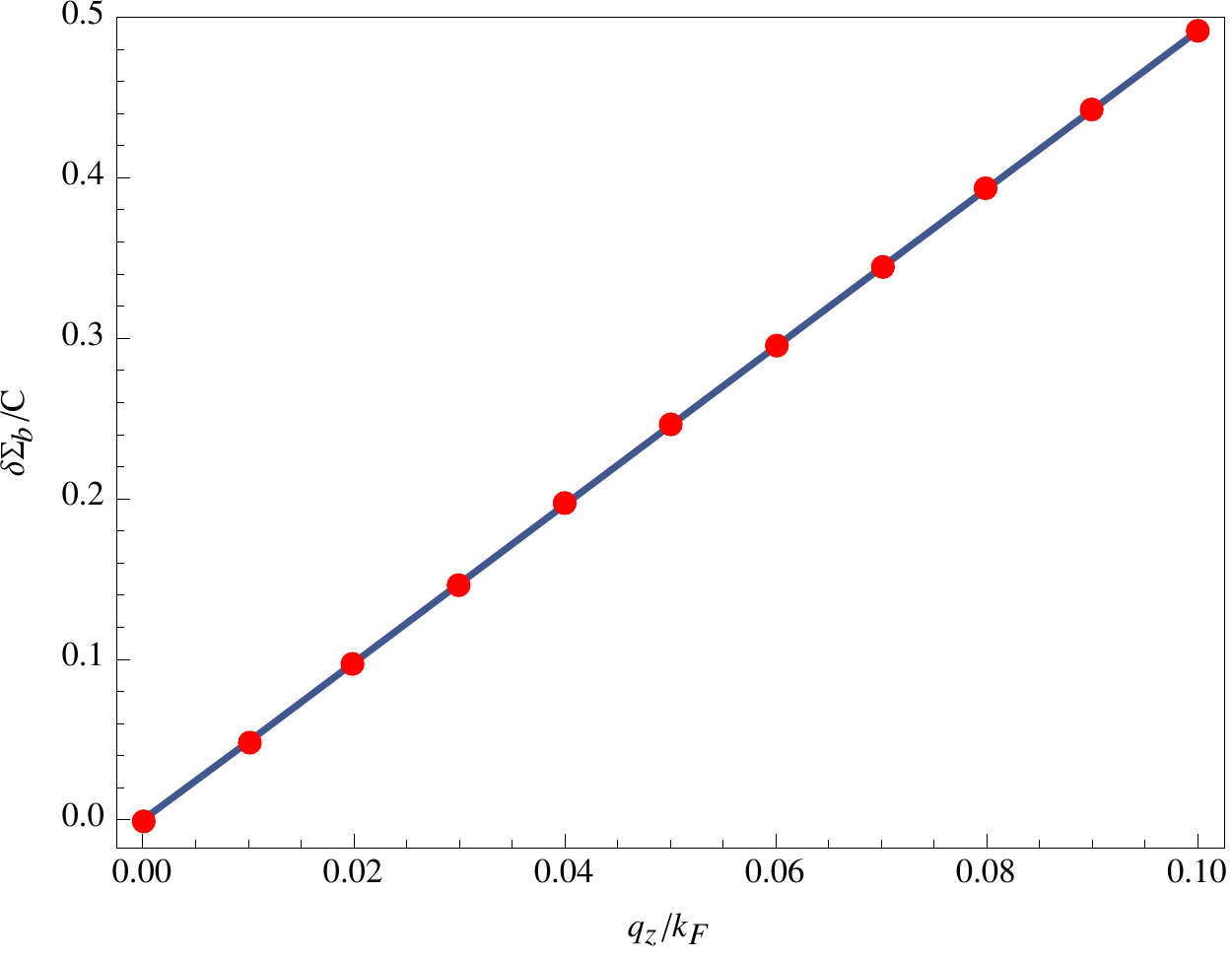}
\caption{Numerical Calculation of Boson Self-energy $\Sigma_b(\bm{q} = q_z \hat{z}, 0)$ normalized by $C=\frac{g^{2}}{(2\pi)^{3}}\frac{2\pi m}{v_{z}}$. Here we set $k_{F} = 1$, $m=\frac{1}{2}$ and $v_{z}=1$ for the numerical calculation. This clearly demonstrates $ \Sigma_b(\bm{q} = q_z \hat{z}, 0) \sim |q_z|$. 
 \label{qz}
}
\end{center}
\end{figure}

We start with the following 
\begin{align}
\notag\Sigma_{b}(0,q_{z})=&\frac{g^{2}}{(2\pi)^{3}}\int d^{3}k\text{Tr}[\text{P}_{\alpha}(\mathbf{k}+\mathbf{q}/2)\tau^{y}\text{P}_{\gamma}(\mathbf{k}-\mathbf{q}/2)\tau^{y}]\frac{n_{F}(\alpha E(\mathbf{k}+\mathbf{q}/2))-n_{F}(\gamma E(\mathbf{k}-\mathbf{q}/2))}{\alpha E(\mathbf{k}+\mathbf{q}/2)-\gamma E(\mathbf{k}-\mathbf{q}/2)},\\
\notag=&\frac{2g^{2}}{(2\pi)^{3}}\int d^{3}k\left\{\text{Tr}[\text{P}_{+}(\mathbf{k}+\mathbf{q}/2)\tau^{y}\text{P}_{-}(\mathbf{k}-\mathbf{q}/2)\tau^{y}]\frac{n_{F}( E(\mathbf{k}+\mathbf{q}/2))+n_{F}(E(\mathbf{k}-\mathbf{q}/2))-1}{E(\mathbf{k}+\mathbf{q}/2)+E(\mathbf{k}-\mathbf{q}/2)}\right.\\
&\left.\quad\quad\quad\quad\quad\quad+\text{Tr}[\text{P}_{+}(\mathbf{k}+\mathbf{q}/2)\tau^{y}\text{P}_{+}(\mathbf{k}-\mathbf{q}/2)\tau^{y}]\frac{n_{F}( E(\mathbf{k}+\mathbf{q}/2))-n_{F}(E(\mathbf{k}-\mathbf{q}/2))}{E(\mathbf{k}+\mathbf{q}/2)-E(\mathbf{k}-\mathbf{q}/2)}\right\},
\end{align}
where
\begin{equation}
\begin{aligned}
\text{Tr}[\text{P}_{\pm}(\mathbf{k}+\mathbf{q}/2)\tau^{y}\text{P}_{\pm}(\mathbf{k}-\mathbf{q}/2)\tau^{y}]=&\frac{1}{2}\left(1-\frac{\epsilon_{x}(\mathbf{k})^{2}+\epsilon_{z}(\mathbf{k}+\mathbf{q}_{2}/2)\epsilon_{z}(\mathbf{k}-\mathbf{q}_{2}/2)}{E(\mathbf{k}+\mathbf{q}/2)E(\mathbf{k}-\mathbf{q}/2)}\right),\\
\text{Tr}[\text{P}_{\pm}(\mathbf{k}+\mathbf{q}/2)\tau^{y}\text{P}_{\mp}(\mathbf{k}-\mathbf{q}/2)\tau^{y}]=&\frac{1}{2}\left(1+\frac{\epsilon_{x}(\mathbf{k})^{2}+\epsilon_{z}(\mathbf{k}+\mathbf{q}_{2}/2)\epsilon_{z}(\mathbf{k}-\mathbf{q}_{2}/2)}{E(\mathbf{k}+\mathbf{q}/2)E(\mathbf{k}-\mathbf{q}/2)}\right). 
\end{aligned}
\end{equation}
Being interested in the quantum critical dynamics, we set the temperature to be zero, i.e., $\beta\rightarrow\infty$, to find 
\begin{align}
\notag\Sigma_{b}(0,q_{z})=&-\frac{g^{2}}{(2\pi)^{3}}\int d^{3}k\left(1+\frac{\epsilon_{x}(\mathbf{k})^{2}+\epsilon_{z}(\mathbf{k}+\mathbf{q}_{2}/2)\epsilon_{z}(\mathbf{k}-\mathbf{q}_{2}/2)}{E(\mathbf{k}+\mathbf{q}/2)E(\mathbf{k}-\mathbf{q}/2)}\right)\frac{1}{E(\mathbf{k}+\mathbf{q}/2)+E(\mathbf{k}-\mathbf{q}/2)},\\
\notag=&-\frac{g^{2}}{(2\pi)^{3}}\frac{2\pi m}{v_{z}}\int_{0}^{\infty}d\epsilon_{\perp}\int_{-\infty}^{\infty}d\epsilon_{z}\left(1+\frac{(\epsilon_{\perp}-\epsilon_{F})^{2}+(\epsilon_{z}+v_{z}q_{z}/2)(\epsilon_{z}-v_{z}q_{z}/2)}{\sqrt{(\epsilon_{\perp}-\epsilon_{F})^{2}+(\epsilon_{z}+v_{z}q_{z}/2)^{2}}\sqrt{(\epsilon_{\perp}-\epsilon_{F})^{2}+(\epsilon_{z}-v_{z}q_{z}/2)^{2}}}\right),\\
&\quad\quad\quad\quad\times\frac{1}{\sqrt{(\epsilon_{\perp}-\epsilon_{F})^{2}+(\epsilon_{z}+v_{z}q_{z}/2)^{2}}+\sqrt{(\epsilon_{\perp}-\epsilon_{F})^{2}+(\epsilon_{z}-v_{z}q_{z}/2)^{2}}},
\end{align}
where $\epsilon_{\perp}=(k_{x}^{2}+k_{y}^{2})/2m$, $\epsilon_{F}=k_{F}^{2}/2m$ and $\epsilon_{z}=v_{z}k_{z}$. As before, we shift $\epsilon_\perp$ to find  \begin{align}
\notag\Sigma_{b}(0,q_{z})=&-\frac{g^{2}}{(2\pi)^{3}}\frac{2\pi m}{v_{z}}\int_{-\epsilon_{F}}^{\infty}d\epsilon_{\perp}\int_{-\infty}^{\infty}d\epsilon_{z}\left(1+\frac{\epsilon_{\perp}^{2}+(\epsilon_{z}+v_{z}q_{z}/2)(\epsilon_{z}-v_{z}q_{z}/2)}{\sqrt{\epsilon_{\perp}^{2}+(\epsilon_{z}+v_{z}q_{z}/2)^{2}}\sqrt{\epsilon_{\perp}^{2}+(\epsilon_{z}-v_{z}q_{z}/2)^{2}}}\right)\\
&\quad\quad\quad\quad\times\frac{1}{\sqrt{\epsilon_{\perp}^{2}+(\epsilon_{z}+v_{z}q_{z}/2)^{2}}+\sqrt{\epsilon_{\perp}^{2}+(\epsilon_{z}-v_{z}q_{z}/2)^{2}}}.
\end{align}
As before, this has the divergent part which we subtract out and we concentrate on the finite contribution
\begin{align}
\notag\Sigma_{b}(0,q_{z})=&-\frac{g^{2}}{(2\pi)^{3}}\frac{2\pi m}{v_{z}}\int_{-\epsilon_{F}}^{\infty}d\epsilon_{\perp}\int_{-\infty}^{\infty}d\epsilon_{z}\left[\left(1+\frac{\epsilon_{\perp}^{2}+(\epsilon_{z}+v_{z}q_{z}/2)(\epsilon_{z}-v_{z}q_{z}/2)}{\sqrt{\epsilon_{\perp}^{2}+(\epsilon_{z}+v_{z}q_{z}/2)^{2}}\sqrt{\epsilon_{\perp}^{2}+(\epsilon_{z}-v_{z}q_{z}/2)^{2}}}\right)\right.\\
&\left.\quad\quad\quad\quad\times\frac{1}{\sqrt{\epsilon_{\perp}^{2}+(\epsilon_{z}+v_{z}q_{z}/2)^{2}}+\sqrt{\epsilon_{\perp}^{2}+(\epsilon_{z}-v_{z}q_{z}/2)^{2}}} +\frac{1}{\sqrt{\epsilon_{\perp}^{2}+\epsilon_{z}^{2}}}\right].
\end{align}
Next we scale the variables, $\epsilon_{\perp}=v_{z}|q_{z}|x$ and $\epsilon_{z}=v_{z}|q_{z}|z$, and find 
\begin{align}
\notag\delta\Sigma_{b}(0,q_{z})=&-\frac{g^{2}}{(2\pi)^{3}}{2\pi m}|q_{z}|\int_{-\epsilon_{F}/v_{z}|q_{z}|}^{\infty}dx\int_{-\infty}^{\infty}dz\left[\left(1+\frac{x^{2}+(z+1/2)(z-1/2)}{\sqrt{x^{2}+(z+1/2)^{2}}\sqrt{x^{2}+(z-1/2)^{2}}}\right)\right.\\
&\left.\quad\quad\quad\quad\times\frac{1}{\sqrt{x^{2}+(z+1/2)^{2}}+\sqrt{x^{2}+(z-1/2)^{2}}}-\frac{1}{\sqrt{x^{2}+z^{2}}}\right], 
\end{align}
For the long-distance limit, i.e., $\epsilon_{F}\gg|q_{z}|$, we approximate $-\epsilon_{F}/|\Omega_{n}|\to -\infty$
\begin{align}
\notag\delta\Sigma_{b}(0,q_{z})=&-\frac{g^{2}}{(2\pi)^{3}}{2\pi m}|q_{z}|\int_{-\infty}^{\infty}dx\int_{-\infty}^{\infty}dz\left[\left(1+\frac{x^{2}+(z+1/2)(z-1/2)}{\sqrt{x^{2}+(z+1/2)^{2}}\sqrt{x^{2}+(z-1/2)^{2}}}\right)\right.\\
\notag&\left.\quad\quad\quad\quad\times\frac{1}{\sqrt{x^{2}+(z+1/2)^{2}}+\sqrt{x^{2}+(z-1/2)^{2}}}-\frac{1}{\sqrt{x^{2}+z^{2}}}\right],\\
=&\frac{g^{2}}{(2\pi)^{3}}\frac{2\pi m}{v_{z}}v_{z}|q_{z}|C_{z},
\end{align}
where
\begin{align}
\notag C_{z}=&\int_{-\infty}^{\infty}dx\int_{-\infty}^{\infty}dz\left[\frac{1}{\sqrt{x^{2}+z^{2}}}-\left(1+\frac{x^{2}+(z+1/2)(z-1/2)}{\sqrt{x^{2}+(z+1/2)^{2}}\sqrt{x^{2}+(z-1/2)^{2}}}\right)\frac{1}{\sqrt{x^{2}+(z+1/2)^{2}}+\sqrt{x^{2}+(z-1/2)^{2}}}\right],\\
=&4.90726. 
\end{align}
This calculation confirms the expected behavior $\Sigma_b (\bm{q}, 0) \sim |q_z|$ with $\bm{q} = q_z \hat{z}$. We can also perform a numerical integral to confirm this behavior in Fig. \ref{qz}.  

\textbf{c. $q_{\perp}$ Dependence}: Next we calculate the dependence of the boson self-energy on $q_\perp$. To see this, we set  $\bm{q}=(q_x,0,0)$ with $\Omega =0$. We expect to see 
\begin{align}
\Sigma_b (\bm{q}, 0) \sim |q_x| = |q_\perp|,  \nonumber
\end{align}
because of the rotational symmetry in xy-plane. 

\begin{figure}[h]
\begin{center}
\includegraphics[width=0.5\textwidth]{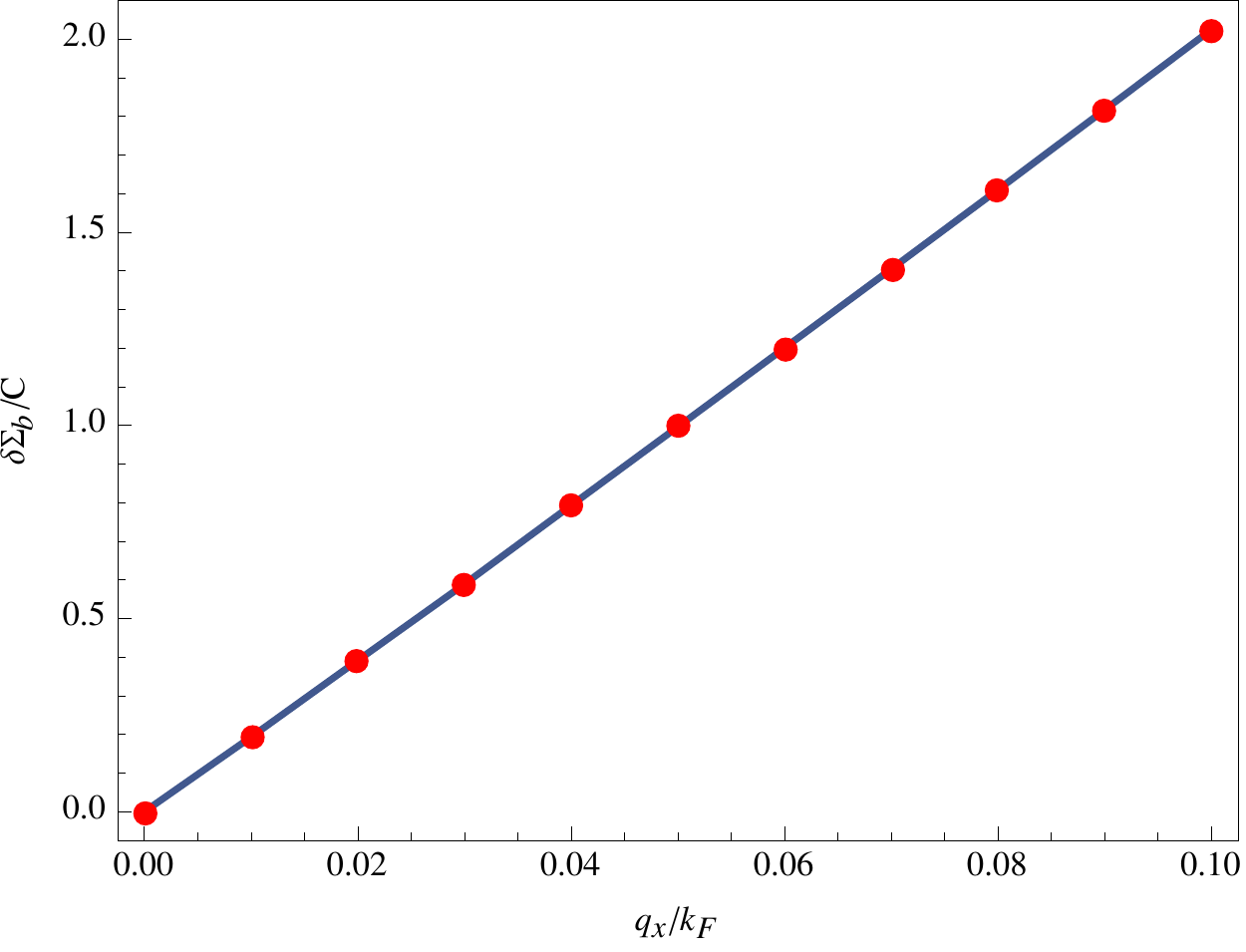}
\caption{Numerical Calculation of Boson Self-energy $\Sigma_b(\bm{q} = q_\perp \hat{x}, 0)$ normalized by $C=\frac{g^{2}}{(2\pi)^{3}}\frac{1}{v_{z}}$. Here we set $k_{F} = 1$, $m=\frac{1}{2}$ and $v_{z}=1$ for the numerical calculation. This clearly demonstrates $ \Sigma_b(\bm{q} = q_\perp \hat{x}, 0) \sim |q_\perp|$. 
 \label{qx}
}
\end{center}
\end{figure}

We start with the following expression of the self-energy 
\begin{align}
\notag\Sigma_{b}(0,q_{x})=&\frac{g^{2}}{(2\pi)^{3}}\int d^{3}k\text{Tr}[\text{P}_{\alpha}(\mathbf{k}+\mathbf{q}/2)\tau^{y}\text{P}_{\gamma}(\mathbf{k}-\mathbf{q}/2)\tau^{y}]\frac{n_{F}(\alpha E(\mathbf{k}+\mathbf{q}/2))-n_{F}(\gamma E(\mathbf{k}-\mathbf{q}/2))}{\alpha E(\mathbf{k}+\mathbf{q}/2)-\gamma E(\mathbf{k}-\mathbf{q}/2)},\\
\notag=&\frac{2g^{2}}{(2\pi)^{3}}\int d^{3}k\left\{\text{Tr}[\text{P}_{+}(\mathbf{k}+\mathbf{q}/2)\tau^{y}\text{P}_{-}(\mathbf{k}-\mathbf{q}/2)\tau^{y}]\frac{n_{F}( E(\mathbf{k}+\mathbf{q}/2))+n_{F}(E(\mathbf{k}-\mathbf{q}/2))-1}{E(\mathbf{k}+\mathbf{q}/2)+E(\mathbf{k}-\mathbf{q}/2)}\right.\\
&\left.\quad\quad\quad\quad\quad\quad+\text{Tr}[\text{P}_{+}(\mathbf{k}+\mathbf{q}/2)\tau^{y}\text{P}_{+}(\mathbf{k}-\mathbf{q}/2)\tau^{y}]\frac{n_{F}( E(\mathbf{k}+\mathbf{q}/2))-n_{F}(E(\mathbf{k}-\mathbf{q}/2))}{E(\mathbf{k}+\mathbf{q}/2)-E(\mathbf{k}-\mathbf{q}/2)}\right\},
\end{align}
where
\begin{equation}
\begin{aligned}
\text{Tr}[\text{P}_{\pm}(\mathbf{k}+\mathbf{q}/2)\tau^{y}\text{P}_{\pm}(\mathbf{k}-\mathbf{q}/2)\tau^{y}]=&\frac{1}{2}\left(1-\frac{\epsilon_{x}(\mathbf{k}+\mathbf{q}/2)\epsilon_{x}(\mathbf{k}-\mathbf{q}/2)+\epsilon_{z}(\mathbf{k})^{2}}{E(\mathbf{k}+\mathbf{q}/2)E(\mathbf{k}-\mathbf{q}/2)}\right),\\
\text{Tr}[\text{P}_{\pm}(\mathbf{k}+\mathbf{q}/2)\tau^{y}\text{P}_{\mp}(\mathbf{k}-\mathbf{q}/2)\tau^{y}]=&\frac{1}{2}\left(1+\frac{\epsilon_{x}(\mathbf{k}+\mathbf{q}/2)\epsilon_{x}(\mathbf{k}-\mathbf{q}/2)+\epsilon_{z}(\mathbf{k})^{2}}{E(\mathbf{k}+\mathbf{q}/2)E(\mathbf{k}-\mathbf{q}/2)}\right). 
\end{aligned}
\end{equation}
At zero temperature, we set $\beta\rightarrow\infty$ to find
\begin{align}
\notag\Sigma_{b}(0,q_{x})=&-\frac{g^{2}}{(2\pi)^{3}}\int d^{3}k\left(1+\frac{\epsilon_{x}(\mathbf{k}+\mathbf{q}/2)\epsilon_{x}(\mathbf{k}-\mathbf{q}/2)+\epsilon_{z}(\mathbf{k})^{2}}{E(\mathbf{k}+\mathbf{q}/2)E(\mathbf{k}-\mathbf{q}/2)}\right)\frac{1}{E(\mathbf{k}+\mathbf{q}/2)+E(\mathbf{k}-\mathbf{q}/2)},\\
\notag=&-\frac{g^{2}}{(2\pi)^{3}}\frac{1}{v_{z}}\int dk_{x}dk_{y}d\epsilon_{z}\left(1+\frac{((k_{x}+q_{x}/2)^{2}+k_{y}^{2}-k_{F}^{2})((k_{x}-q_{x}/2)^{2}+k_{y}^{2}-k_{F}^{2})/4m^{2}+\epsilon_{z}^{2}}{\sqrt{((k_{x}+q_{x}/2)^{2}+k_{y}^{2}-k_{F}^{2})^{2}/4m^{2}+\epsilon_{z}^{2}}\sqrt{((k_{x}-q_{x}/2)^{2}+k_{y}^{2}-k_{F}^{2})^{2}/4m^{2}+\epsilon_{z}^{2}}}\right)\\
&\quad\quad\quad\quad\times\frac{1}{\sqrt{((k_{x}+q_{x}/2)^{2}+k_{y}^{2}-k_{F}^{2})^{2}/4m^{2}+\epsilon_{z}^{2}}+\sqrt{((k_{x}-q_{x}/2)^{2}+k_{y}^{2}-k_{F}^{2})^{2}/4m^{2}+\epsilon_{z}^{2}}},
\end{align}
where $\epsilon_{z}=v_{z}k_{z}$. 
As before, we subtract the divergent part to extract the finite piece 
\begin{align}
\notag\delta\Sigma_{b}(0,q_{x})=&-\frac{g^{2}}{(2\pi)^{3}}\frac{1}{v_{z}}\int dk_{x}dk_{y}d\epsilon_{z}\left[\left(1+\frac{((k_{x}+q_{x}/2)^{2}+k_{y}^{2}-k_{F}^{2})((k_{x}-q_{x}/2)^{2}+k_{y}^{2}-k_{F}^{2})/4m^{2}+\epsilon_{z}^{2}}{\sqrt{((k_{x}+q_{x}/2)^{2}+k_{y}^{2}-k_{F}^{2})^{2}/4m^{2}+\epsilon_{z}^{2}}\sqrt{((k_{x}-q_{x}/2)^{2}+k_{y}^{2}-k_{F}^{2})^{2}/4m^{2}+\epsilon_{z}^{2}}}\right)\right.\\
\notag&\quad\quad\quad\quad\quad\quad\quad\times\frac{1}{\sqrt{((k_{x}+q_{x}/2)^{2}+k_{y}^{2}-k_{F}^{2})^{2}/4m^{2}+\epsilon_{z}^{2}}+\sqrt{((k_{x}-q_{x}/2)^{2}+k_{y}^{2}-k_{F}^{2})^{2}/4m^{2}+\epsilon_{z}^{2}}}\\
&\quad\quad\quad\quad\quad\quad\quad\left.-\frac{1}{\sqrt{(k_{x}^{2}+k_{y}^{2}-k_{F}^{2})^{2}/4m^{2}+\epsilon_{z}^{2}}}\right], 
\end{align}
which we can now evaluate numerically. From the numerical calculation, Fig. \ref{qx}, we clearly see the expected behavior $\Sigma_b(\bm{q} = q_x \hat{x}, 0) \sim |q_x| = |q_\perp|$. 


\subsection{Fermion Self-Energy} 
Here we present the detailed calculation of the fermion self-energy $\Sigma_f(\bm{k},\omega)$ to the lowest order in $O(\tfrac{1}{N_f})$ expansion. 
\begin{align}
\Sigma_f (\bm{k},\omega) = g^2 \int_{\bm{q},\Omega}\mathcal{F}(\theta_{\bm{k}})\mathcal{F}(\theta_{\bm{k}+\bm{q}}) \tau^y G_{f,0}(\bm{k}+\bm{q}, \omega+\Omega) \tau^y G_b(\bm{q},\Omega), 
\end{align}
in which $(\bm{q}, \Omega)$ is the bosonic momentum and frequency. We also have used the renormalized boson propagator $G_{b}(\bm{q}, \Omega) \approx \frac{1}{\Sigma_b(\bm{q},\Omega)}$. Using the approximation $\mathcal{F}(\theta_{\bm{k} + \bm{q}}) \approx \mathcal{F}(\theta_{\bm{k}})$, we have 
\begin{align}
\Sigma_f (\bm{k},\omega) \propto g^2 \mathcal{F}^2 (\theta_{\bm{k}}) \int_{\bm{q},\Omega} \tau^y G_{f,0}(\bm{k}+\bm{q}, \omega+\Omega) \tau^y G_b(\bm{q},\Omega), 
\end{align}
We will show that the fermion self-energy has no IR divergence, and this requires us to identify the cutoffs explicitly in the integrals. We use the following cutoff scheme for the integral, 
\begin{align}
\int_{\bm{q}, \Omega} = \Big( \int^{\Lambda}_\mu \frac{q_\perp dq_\perp}{2\pi} + \int^{-\mu}_{-\Lambda} \frac{q_\perp dq_\perp}{2\pi} \Big)\int^{2\pi}_0 \frac{d\phi_{\bm{q}}}{2\pi} \int^{\infty}_{-\infty} \frac{dq_z}{2\pi} \int^{\infty}_{-\infty} \frac{d\Omega}{2\pi},  
\label{BosonIntegral}
\end{align} 
and we will show that $\Sigma_f(\bm{k},\omega)$ has no divergence in $\mu \to 0^{+}$. 

We first start with the fermion propagator $G_{f,0}(\bm{k}+\bm{q},\omega+\Omega)$ in which 
\begin{align}
G_{f,0}(\bm{k}+\bm{q}, \omega+\Omega) &= \frac{1}{-i(\omega+\Omega) + v_\perp (|\bm{k}_\perp + \bm{q}_\perp| - k_f)\tau^x + v_z (\delta k_z + q_z)\tau^z},  \nonumber\\ 
&\equiv \frac{1}{-i\epsilon_0 + \epsilon_x \tau^x + \epsilon_z \tau^z}, 
\end{align}
in which 
\begin{align}
&\epsilon_0 = \omega + \Omega, \nonumber\\ 
&\epsilon_x = v_\perp (|\bm{k}_\perp + \bm{q}_\perp| - k_f), \nonumber\\ 
&\epsilon_z = v_z (\delta k_z + q_z). \nonumber
\end{align} 
Using this representation, we find 
\begin{align}
\Sigma_f (\bm{k}, \omega) \propto g^2 \mathcal{F}^2(\theta_{\bm{k}}) \int_{\bm{q}, \Omega}  \frac{-i\epsilon_0 + \epsilon_x \tau^x + \epsilon_z \tau^z}{\epsilon_0^2 + \epsilon_x^2 + \epsilon_z^2} G_b(\bm{q}, \Omega). 
\end{align}
To evaluate the integrals, we first note that 
\begin{align}
|\bm{k}_\perp + \bm{q}_\perp| \approx k_f + \delta k_\perp + q_\perp \cos(\phi_{\bm{q}}),  
\end{align}
in which we have used $|\bm{k}| = k_f + \delta k_\perp$ and $|\delta k_\perp|, q_\perp \ll k_f$, and $\phi_{\bm{q}}$ is the angle between $\bm{k}_\perp$ and $\bm{q}_\perp$. Using this, we find 
\begin{align}
\Sigma_f (\bm{k}, \omega) \propto g^2 \mathcal{F}^2(\theta_{\bm{k}}) \int_{\bm{q}, \Omega}  \frac{-i(\omega+\Omega)+ v_\perp(\delta k_\perp + q_\perp \cos(\phi_{\bm{q}})) \tau^x + v_z(q_z + \delta k_z) \tau^z}{(\omega+\Omega)^2 + v_\perp^2 (\delta k_\perp + q_\perp \cos(\phi_{\bm{q}}))^2 + v_z^2 (q_z + \delta k_z)^2} G_b(\bm{q}, \Omega). 
\label{FermionSelfE}
\end{align}
To find the renormalization of $G_{f,0}^{-1} (\bm{k}, \omega) = -i\omega + v_z \delta k_z \tau^z \mu^z + v_\perp \delta k_\perp \tau^x$, we need to expand the fermion self-energy $-\Sigma_f(\bm{k},\omega)$ Eq.\eqref{FermionSelfE} for small $\delta \bm{k}$ and $\omega$ to the lowest orders. 

The lowest order is of couse $-\Sigma_f (\bm{k}^{*}, 0)$ in which $\bm{k}^{*}$ is exactly on the nodal ring, i.e., $\delta \bm{k}_\perp = \bm{0}$ and $\delta k_z = 0$. For such momentum and frequency, the fermion self-energy is 
\begin{align}
-\Sigma_f (\bm{k}^{*}, 0) \propto g^2 \mathcal{F}^2(\theta_{\bm{k}}) \int_{\bm{q}, \Omega}  \Big[-i\Omega+ v_\perp q_\perp \cos(\phi_{\bm{q}}) \tau^x + v_z q_z  \tau^z \Big]\times \frac{G_b(\bm{q}, \Omega)}{\Omega^2 + v_\perp^2  q_\perp^2 \cos^2(\phi_{\bm{q}}) + v_z^2 q_z^2}, 
\end{align}
which we will show to be zero. (Note that this is correct up to $O(|\bm{q}|/k_F)$ because of the approximation $\mathcal{F}(\theta_{\bm{k}+\bm{q}}) = \mathcal{F}(\theta_{\bm{k}}) + O(|\bm{q}|/k_F) \approx \mathcal{F}(\theta_{\bm{k}})$). To see this, we note that 
\begin{align}
&O(\Omega, q_z, q_\perp, \phi_{\bm{q}}) \equiv \Big[-i\Omega+ v_\perp q_\perp \cos(\phi_{\bm{q}}) \tau^x + v_z q_z  \tau^z \Big],\nonumber\\ 
&E(\Omega, q_z, q_\perp, \phi_{\bm{q}}) \equiv \frac{G_b(\bm{q}, \Omega)}{\Omega^2 + v_\perp^2  q_\perp^2 \cos^2(\phi_{\bm{q}}) + v_z^2 q_z^2}, \nonumber\\ 
\end{align}
which are appearing in the fermion self-energy via 
\begin{align}
&-\Sigma_f (\bm{k}^{*}, 0) = g^2 \mathcal{F}^2(\theta_{\bm{k}}) \int_{\bm{q}, \Omega}  E(\Omega, q_z, q_\perp, \phi_{\bm{q}}) \cdot O(\Omega, q_z, q_\perp, \phi_{\bm{q}}). 
\end{align}
Now by noting that $O(\cdots)$ is an odd function in $\bm{q}$ and $E(\cdots)$ is an even function in $\bm{q}$, it is straightforward to see $-\Sigma_f (\bm{k}^{*}, 0) = 0$. This implies, as explained in the main text, that there is no correction to the size and positions of the nodal ring due to the critical boson within the approximation $\mathcal{F}(\theta_{\bm{k}+\bm{q}}) = \mathcal{F}(\theta_{\bm{k}}) + O(|\bm{q}|/k_F) \approx \mathcal{F}(\theta_{\bm{k}})$. 
  
On confirming $-\Sigma_f (\bm{k}^{*}, 0) = 0$, we consider the higher order terms, which are the corrections $\{ J_0, J_1, J_2 \}$ to the fermion propagator  
\begin{align}
G_{f}^{-1}(\bm{k}, \omega) = G_{f,0}^{-1}(\bm{k},\omega) - \Sigma_f(\bm{k},\omega) \to G_{f}^{-1}(\bm{k},\omega) = -i\omega (1+ J_0) + v_z(1+ J_1) \delta k_z \tau^z \mu^z + v_\perp(1+ J_2) \delta k_\perp \tau^x, 
\end{align}
by performing the expansion for $\Sigma_f(\bm{k},\omega)$ for small $\delta \bm{k}$ and $\omega$. The integrals $\{J_0, J_1, J_2 \}$ are the following. 
\begin{align}
&J_0 = g^2 \mathcal{F}^2(\theta_{\bm{k}})\int_{\bm{q}, \Omega} G_b(\bm{q}, \Omega) \frac{-\Omega^2 + v_\perp^2  q_\perp^2 \cos^2(\phi_{\bm{q}}) + v_z^2 q_z^2}{(\Omega^2 + v_\perp^2  q_\perp^2 \cos^2(\phi_{\bm{q}}) + v_z^2 q_z^2)^2}, \nonumber\\ 
& J_1 = g^2 \mathcal{F}^2(\theta_{\bm{k}})\int_{\bm{q}, \Omega} G_b(\bm{q}, \Omega) \frac{\Omega^2 + v_\perp^2  q_\perp^2 \cos^2(\phi_{\bm{q}}) - v_z^2 q_z^2}{(\Omega^2 + v_\perp^2  q_\perp^2 \cos^2(\phi_{\bm{q}}) + v_z^2 q_z^2)^2},
\nonumber\\ 
& J_2 = g^2 \mathcal{F}^2(\theta_{\bm{k}})\int_{\bm{q}, \Omega} G_b(\bm{q}, \Omega) \frac{\Omega^2 - v_\perp^2  q_\perp^2 \cos^2(\phi_{\bm{q}}) + v_z^2 q_z^2}{(\Omega^2 + v_\perp^2  q_\perp^2 \cos^2(\phi_{\bm{q}}) + v_z^2 q_z^2)^2}. 
\end{align}
The three integrals can be evaluated in exactly the same fashion and so we present the detailed calculation only for $J_2$. We first write out $J_2$ explicitly  
\begin{align}
J_2 &\propto \frac{\mathcal{F}^2(\theta_{\bm{k}})}{N_f} \int q_\perp dq_\perp \int dq_z \int d\phi_{\bm{q}} \int d\Omega \frac{1}{\int^{2\pi}_0 d\chi \mathcal{F}^2(\chi) \sqrt{\Omega^2 + v_z^2 q_z^2 + v_\perp^2 q_\perp^2 \cos^2 (\chi)} } \frac{\Omega^2 - v_\perp^2  q_\perp^2 \cos^2(\phi_{\bm{q}}) + v_z^2 q_z^2}{(\Omega^2 + v_\perp^2  q_\perp^2 \cos^2(\phi_{\bm{q}}) + v_z^2 q_z^2)^2}, \nonumber
\end{align}  
in which the integral is apparently divergent without the cutoffs. Hence we introduce the cutoffs over the momentum and frequency by following the cutoff scheme Eq.\eqref{BosonIntegral}. Now we perform the change of the variables as following $q_z = \frac{v_\perp q_\perp}{v_z} a$ and $\Omega = v_\perp q_\perp b$. Then the integral $J_2$ becomes 
\begin{align}
&J_2 \propto\frac{\mathcal{F}^2(\theta_{\bm{k}})}{2N_f} \Big( \int^{\Lambda}_\mu dq_\perp \Big) \times I_\perp = \frac{ \mathcal{F}^2(\theta_{\bm{k}})}{2N_f}\Big(\Lambda - \mu \Big)  I_\perp, 
\end{align}
in which it is safe to bring $\mu \to 0^{+}$ without encountering any singularity. Here the numeric integral $I_\perp$ is following:
\begin{align}
&I_\perp = \int^{\infty}_{-\infty} da \int^{\infty}_{-\infty} db \int^{2\pi}_0 d\phi \frac{a^2 + b^2 - \cos^2(\phi)}{[a^2 + b^2 + \cos^2(\phi)]^2} \frac{1}{\int^{2\pi}_0 d\chi \mathcal{F}^2(\chi)\sqrt{a^2 + b^2 + \cos^2(\chi)}}. 
\end{align}
This integral is well-defined and finite because 
\begin{align}
I_\perp &= \int^{\infty}_{-\infty} da \int^{\infty}_{-\infty} db \int^{2\pi}_0 d\phi \frac{a^2 + b^2 - \cos^2(\phi)}{[a^2 + b^2 + \cos^2(\phi)]^2} \frac{1}{\int^{2\pi}_0 d\chi \mathcal{F}^2(\chi)\sqrt{a^2 + b^2 + \cos^2(\chi)}} \nonumber\\ 
&\propto \int d\phi \int dr \frac{r(r^2 - \cos^2(\phi))}{(r^2 + \cos^2 (\phi))^2} \frac{1}{\int^{2\pi}_0 d\chi \sqrt{r^2 + \cos^2(\chi)}} \equiv \int d\phi \int dr H[r, \phi],
\end{align}
where the integral over $\phi$ cannot have any singularity as $H[r,\phi]$ is always regular in $\phi$. On the other hand, the integral over $r$ is also regular because $H[r \to 0, \phi] \to r$ and $H[r\to \infty, \phi] \to \frac{1}{r^2}$. Hence we conclude that 
\begin{align}
J_2 \propto  \frac{ \mathcal{F}^2(\theta_{\bm{k}})}{2N_f} \Big(\Lambda - \mu \Big), 
\end{align}
which has no IR divergence. Similarly, we can show that
\begin{align}
J_i \propto  \frac{ \mathcal{F}^2(\theta_{\bm{k}})}{2N_f} \Big(\Lambda - \mu \Big), ~i=0,1,2, 
\end{align}
and hence there is no IR divergence in the fermion self-energy. 

\subsection{Vertex Correction}
Here we compute the vertex correction with the zero fermionic incoming momentum and frequency, $\delta k_z = \delta k_\perp =0$ and $\omega = 0$, i.e., the fermion mode is exactly on the nodal line (see Fig. 2 of the main text). Hence, we only specify the azimuthal angle $\theta_{\bm{k}}$ of the fermionic momentum $\bm{k}$. Now the vertex correction $\tau^y \to \tau^y + \Gamma^y$ is given by  
\begin{align}
\Gamma^y \propto \int_{\bm{q}, \Omega} g^3 \mathcal{F}^3 (\theta_{\bm{k}}) \tau^y G_f (\Omega, \bm{k}+\bm{q}) \tau^y G_f (\Omega, \bm{k}+\bm{q}) \tau^y G_b (\Omega, \bm{q}),
\end{align}
in which $\bm{q}$ is the bosonic momentum centered near $\bm{q} \approx 0$ (see Fig. 2 of the main text). We now use the linearized dispersion for the fermions $|\bm{k}_\perp + \bm{q}_\perp| \approx k_f + q_\perp \cos(\phi_{\bm{q}})$ to find 
\begin{align}
\Gamma^y & \propto g^3\mathcal{F}^3 (\theta_{\bm{k}}) \tau^y \int_{\bm{q}, \Omega} \frac{1}{\Omega^2 + v_z^2 q_z^2 + v_\perp^2 q_\perp^2 \cos^2(\phi_{\bm{q}})} G_b (\Omega, \bm{q}) \nonumber\\ 
&\propto \frac{g}{N_f} \mathcal{F}^3 (\theta_{\bm{k}}) \tau^y \int q_\perp d q_\perp d\phi_{\bm{q}} d q_z d \Omega \frac{1}{\Omega^2 + v_z^2 q_z^2 + v_\perp^2 q_\perp^2 \cos^2(\phi_{\bm{q}})} \frac{1}{\int^{2\pi}_0 d\chi \mathcal{F}^2(\chi) \sqrt{\Omega^2 + v_z^2 q_z^2 + v_\perp^2 q_\perp^2 \cos^2 (\chi)}}.
\end{align}
By performing the change of the variables, $q_z = \frac{v_\perp q_\perp}{v_z} a$ and $\Omega = v_\perp q_\perp b$ and cutting-off the integral over $q_\perp \in \{ \mu, \Lambda \}$, we obtain 
\begin{align}
\Gamma^y & \propto \frac{g}{N_f} \mathcal{F}^3 (\theta_{\bm{k}}) \tau^y I_\Gamma \times \Big(\Lambda - \mu \Big), 
\end{align}
in which $I_\Gamma$ is the regular integral without any divergence  
\begin{align}
I_\Gamma \propto \int d \phi_{\bm{q}} da db \frac{1}{a^2 + b^2 + \cos^2(\phi_{\bm{q}})} \frac{1}{\int^{2\pi}_0 d\chi \mathcal{F}^2(\chi) \sqrt{a^2 + b^2 + \cos^2(\chi)}}. 
\end{align}

Hence we clearly see that the vertex correction $\Gamma^y$ does not have any singularity in sending $\mu \to 0^{+}$, showing that the coupling between the fermion and the order parameter becomes irrelevant at the QCP. 

\subsection{Comment on two-dimensional $E$-representation}
For the two-dimensional $E$-represenation, the theory $\mathcal{S} = \mathcal{S}_\psi +  \mathcal{S}_\phi + \mathcal{S}_{\text{int}}$ should be properly modified to reflect the symmetries and two-dimensional nature of the representation. First of all, the free fermion theory part $\mathcal{S}_\psi$ remains the same form. However, the coupling $\mathcal{S}_{\text{int}}$ between the fermions and bosons should be modified as the following, 
\begin{align}
\mathcal{S}_{\text{int}} = g \Big[ \int_{\bm{q}, \Omega} \phi_{x, \bm{q}, \Omega} \int_{\bm{k},\omega} \mathcal{F}_x (\theta_{\bm{k}}) \Psi^{\dagger}_{\bm{k}+\bm{q},\omega+\Omega} \tau^y \Psi_{\bm{k},\omega}\Big] +  g \Big[ \int_{\bm{q}, \Omega} \phi_{y, \bm{q}, \Omega} \int_{\bm{k},\omega} \mathcal{F}_y(\theta_{\bm{k}}) \Psi^{\dagger}_{\bm{k}+\bm{q},\omega+\Omega} \tau^y \Psi_{\bm{k},\omega}\Big],
\end{align}
 in which we note the doublets of the bosons $\{ \phi_x, \phi_y \}$ and form factors $\{ \mathcal{F}_x(\cdot), \mathcal{F}_y (\cdot) \}$. On the other hand, the bosonic part $\mathcal{S}_\phi$ of the action (up to quadratic orders in the doublet fields) is now 
\begin{align}
\mathcal{S}_{\phi} = \int_{\bm{k},\omega} \Big(\frac{1}{2}(u_1^2 k_x^2 + u_2^2 k_y^2+ \omega^2 + r) |\phi_{x, \bm{k}, \omega}|^2+\frac{1}{2}(u_2^2 k_x^2 + u_1^2 k_y^2+ \omega^2 + r) |\phi_{y, \bm{k}, \omega}|^2 \Big),
\label{bareE}
\end{align}
in which the bosons $\{ \phi_x, \phi_y \}$ may have the anisotropic dispersions along $\hat{x}$- and $\hat{y}$-directions.

With this critical theory, one can proceed as the one-dimensional cases. Following the calculations, we note the important identity  
\begin{align}
\int^{2\pi}_0 d\theta_{\bm{k}} \mathcal{F}_x (\theta_{\bm{k}}) \mathcal{F}_y(\theta_{\bm{k}}) = 0, 
\end{align} 
which implies that the one-loop self-energy corrections to the dynamics of $\phi_x$ and $\phi_y$ are decoupled effectively, i.e., to the leading term in $O(\frac{1}{N_f})$ expansion, we effectively have 
\begin{align}
[\Sigma_b (\bm{q}, \Omega)]^{ij} = \delta^{ij} \times \frac{g^2 k_f N_f}{16 \pi v_\perp v_z} \int^{2\pi}_{0} d\theta \Big(\mathcal{F}(\theta)\Big)^2 \Big[\Omega^2 + v_z^2 q_z^2 + v_\perp^2 q_\perp^2 \cos^2 (\theta)\Big]^{1/2}, 
\end{align} 
which dominates the bare dispersion Eq.\eqref{bareE}. Hence the boson dispersion at the criticality becomes isotropic. Now it is straightforward to see that the fermion self-energy will be of the same form as the one-dimensional represenations because the boson propagator is diagonal, i.e., $[G_b^{-1}]^{ij} = [G_{b,0}^{-1}]^{ij} + [\Sigma_b ]^{ij} \to  [\Sigma_b ]^{ij} \propto \delta^{ij}$, with the same form of the scaling behaviors as in the one-dimensional representations. Hence, the nature of the critical theory remains the same even in the two-dimensional $E$-representation. 

\section{Contribution of Boson to Specific Heat}
The effective action for the order parameter at the critical point is
\begin{align}
\mathcal{S}_{\phi}^{c}=\int_{\Omega_{n},\mathbf{q}}C\int_{0}^{2\pi}d\theta\;\left(\mathcal{F}(\theta)\right)^{2}\sqrt{\Omega_{n}^{2}+v_{z}^{2}q_{z}^{2}+v_{\perp}^{2}q_{\perp}^{2}\cos^{2}\theta}\frac{|\phi_{\mathbf{q},\Omega_{n}}|^{2}}{2},
\end{align}
where $C=\frac{g^{2}N_{f}k_{f}}{16 \pi v_{z}v_{\perp}}$. Since the integrand is continuous, by the mean value theorem for integrals, we can find suitable $\theta_{0}\in(0,2\pi) $ which satisfy
\begin{align}
\int_{0}^{2\pi}d\theta\;\left(\mathcal{F}(\theta)\right)^{2}\sqrt{\Omega_{n}^{2}+v_{z}^{2}q_{z}^{2}+v_{\perp}^{2}q_{\perp}^{2}\cos^{2}\theta}=2\pi\left(\mathcal{F}(\theta_{0})\right)^{2}\sqrt{\Omega_{n}^{2}+v_{z}^{2}q_{z}^{2}+v_{\perp}^{2}q_{\perp}^{2}\cos^{2}\theta_{0}}.
\end{align}
\\
Then, we can write
\begin{align}
\mathcal{S}_{\phi}^{c}=\int_{\Omega_{n},\mathbf{q}}2\pi C\left(\mathcal{F}(\theta_{0})\right)^{2}\sqrt{\Omega_{n}^{2}+v_{z}^{2}q_{z}^{2}+\tilde{v}_{\perp}^{2}q_{\perp}^{2}}\frac{|\phi_{\mathbf{q},\Omega_{n}}|^{2}}{2},
\end{align}
where $\tilde{v}_{\perp}:=v_{\perp}\cos\theta_{0}$. Here, $\theta_{0}$ depends on the other variable, $\theta_{0}=\theta_{0}(\Omega_{n},v_{z},q_{z},v_{\perp},q_{\perp})$, but since the difference is small, we can approximate it as fixed value for each representations.\\
The partition function by path integral is
\begin{align}
Z=&\int d\phi_{\mathbf{q},\Omega}e^{-\mathcal{S}_{\phi}^{c}}\notag\\
=&N\prod_{q_{\perp},q_{z}}\prod_{n}\left[{\beta^{2}}(\Omega_{n}^{2}+v_{z}^{2}q_{z}^{2}+\tilde{v}_{\perp}^{2}q_{\perp}^{2})\right]^{-1/4}.
\end{align}
Taking the logarithm and ignoring constant part, we find the free energy (in unit volume) 
\begin{align}
\mathcal{F}=-\frac{T}{V}\ln Z=&\frac{T}{4}\int \frac{d^{2}q_{\perp}dq_{z}}{(2\pi)^{3}}\sum_{n}\ln\beta^{2}(\Omega_{n}^{2}+v_{z}^{2}q_{z}^{2}+\tilde{v}_{\perp}^{2}q_{\perp}^{2}).
\end{align}
Since
\begin{align}
\sum_{n}\ln\beta^{2}(\Omega_{n}^{2}+x^{2})=&\beta x+2\ln(1-e^{-\beta x})+\text{const.}
\end{align}
Ignoring the constant part, we have
\begin{align}
\mathcal{F}=\frac{T}{4}\int\frac{d^{2}q_{\perp}dq_{z}}{(2\pi)^{3}}\left[ \beta\sqrt{v_{z}^{2}q_{z}^{2}+\tilde{v}_{\perp}^{2}q_{\perp}^{2}}+2\ln(1-e^{- \beta\sqrt{v_{z}^{2}q_{z}^{2}+\tilde{v}_{\perp}^{2}q_{\perp}^{2}}})\right].
\end{align}
The first term on the right hand side diverge and it is zero temperature contribution. So, to obtain finite free energy, we subtract the zero temperature contribution and find 
\begin{align}
\notag\delta \mathcal{F}(T)=&\mathcal{F}(T)-\mathcal{F}(0)=\frac{T}{2}\int\frac{d^{2}q_{\perp}dq_{z}}{(2\pi)^{3}}\ln(1-e^{-\beta\sqrt{v_{z}^{2}q_{z}^{2}+\tilde{v}_{\perp}^{2}q_{\perp}^{2}}})\\
\notag=&\frac{T^{4}}{2 \tilde{v}_{\perp}^{2}v_{z}}\int\frac{dk_{x}dk_{y}dk_{z}}{(2\pi)^{3}}\ln(1-e^{-\sqrt{k_{x}^{2}+k_{y}^{2}+k_{z}^{2}}})\\
\notag=&\frac{T^{4}}{4\pi^{2} \tilde{v}_{\perp}^{2}v_{z}}\int_{0}^{\infty}dr\;r^{2}\ln(1-e^{-r})\\
\notag=&-\frac{T^{4}}{4\pi^{2}\tilde{v}_{\perp}^{2}v_{z}}\frac{\pi^{4}}{45}\\
=&-\frac{\pi^{2}}{180\tilde{v}_{\perp}v_{z}}T^{4}.
\end{align}
Thus the contribution of the order parameter to the total specific heat is
\begin{align}
C_{v}=\frac{\partial}{\partial T}\frac{\partial(\beta\delta\mathcal{F})}{\partial\beta}=\frac{\pi^{2}}{15\tilde{v}_{\perp}v_{z}}T^{3}.
\end{align}

\section{Temperature Dependence of Boson Self-energy Correction}
We compute the boson self-energy at the zero momentum and zero frequency.
\begin{align*}
\Sigma_{b}(\Omega_{m},\mathbf{q})=&g^{2}\int_{\mathbf{k},\omega_{n}}(\mathcal{F}(\theta_{\mathbf{k}}))^{2}\text{Tr}\left[ \tau^{y}G_{f,0}(\omega_{n},\mathbf{k})\tau^{y}G_{f,0}(\omega_{n}+\Omega_{m},\mathbf{k}+\mathbf{q}) \right],
\end{align*}
where
\begin{align*}
G_{f,0}(\omega_{n},\mathbf{k})=&\frac{1}{-i\omega_{n}+\mathcal{H}_{0}(\mathbf{k})}=\frac{1}{-i\omega_{n}+\alpha E_{k}}P_{\alpha}(\mathbf{k}),\\
P_{\pm}(\mathbf{k})=&\frac{1}{2}\left(I+\alpha\frac{\mathcal{H}_{0}(\mathbf{k})}{E_{k}}\right),\;\;\;\;\mathcal{H}_{0}(\mathbf{k})=v_{\perp}k_{\perp}\tau_{x}+v_{z}k_{z}\tau_{z},\\
E_{k}=&\sqrt{v_{\perp}^{2}k_{\perp}^{2}+v_{z}^{2}k_{z}^{2}}.
\end{align*}
For $A_{1g}$ representation, $\mathcal{F}(\theta_{k})=1$, then,
\begin{align*}
\Sigma_{b}(0,0)=&\frac{g^{2}k_{F}}{4\pi^{2}}\int dk_{\perp}dk_{z}\text{Tr}\left[\tau^{y}\text{P}_{\alpha}(\mathbf{k})\tau^{y}\text{P}_{\beta}(\mathbf{k})\right]\frac{n_{F}(\alpha E_{k})-n_{F}(\beta E_{k})}{\alpha E_{k}-\beta E_{k}}\\
=&-\frac{g^{2}k_{F}}{4\pi^{2}}\int dk_{\perp}dk_{z}\frac{1}{E_{k}}\tanh\frac{\beta E_{k}}{2}\\
=&-\frac{g^{2}k_{F}}{4\pi v_{\perp}v_{z}}\int dR\;R\frac{1}{R}\tanh\frac{\beta R}{2}=-\frac{g^{2}k_{F}}{4\pi v_{\perp}v_{z}}\int dR\;\tanh\frac{\beta R}{2},
\end{align*}
where
\begin{align*}
\text{Tr}\left[\tau^{y}\text{P}_{\pm}(\mathbf{k})\tau^{y}\text{P}_{\pm}(\mathbf{k})\right]=&0,\\
\text{Tr}\left[\tau^{y}\text{P}_{\pm}(\mathbf{k})\tau^{y}\text{P}_{\mp}(\mathbf{k})\right]=&1.
\end{align*}
Clearly, it has a linear divergence as expected. To obtain a finite result at the critical point, we subtract the zero temperature contribution,
\begin{align*}
\delta \Sigma_{b}(0,0,T)=&\Sigma_{b}(0,0,T)-\Sigma_{b}(0,0,0)=-\frac{g^{2}k_{F}}{4\pi v_{\perp}v_{z}}\int_{0}^{\infty} dR\;\left(\tanh\frac{\beta R}{2}-1\right)\\
=&\frac{g^{2}k_{F}\ln2 }{2\pi v_{\perp}v_{z}}T.
\end{align*}
Thus, at the zero external momentum and frequency limit, the temperature dependence of the boson self-energy correction is $T$-linear.

\end{document}